\documentclass{article}

     \usepackage[nonatbib,final]{neurips_2020}

\usepackage[square,numbers,sort&compress]{natbib}
\usepackage[utf8]{inputenc} 
\usepackage[T1]{fontenc}    
\usepackage{hyperref}       
\usepackage{url}            
\usepackage{booktabs}       
\usepackage{amsfonts}       
\usepackage{nicefrac}       
\usepackage{microtype}      
\usepackage{graphicx}
\usepackage{amsmath}

\usepackage{pdflscape} 
\usepackage{threeparttable}
\usepackage[title]{appendix}
\usepackage{bibunits}

\title{Gibbs Sampling with People}

\author{%
  Peter M. C. Harrison\textsuperscript{*} \\
  Max Planck Institute for Empirical Aesthetics \\
  Frankfurt \\
  \texttt{peter.harrison@ae.mpg.de} \\
  \And 
  Raja Marjieh\textsuperscript{*} \\
  Max Planck Institute for Empirical Aesthetics \\
  Frankfurt \\
  \texttt{raja.marjieh@ae.mpg.de} \\
  \And 
  Federico Adolfi \\
  Max Planck Institute for Empirical Aesthetics \\
  Frankfurt \\
  \texttt{federico.adolfi@ae.mpg.de} \\
  \And 
  Pol van Rijn \\
  Max Planck Institute for Empirical Aesthetics \\
  Frankfurt \\
  \texttt{pol.van-rijn@ae.mpg.de} \\
  \And 
  Manuel Anglada-Tort \\
  Max Planck Institute for Empirical Aesthetics \\
  Frankfurt \\
  \texttt{manuel.anglada-tort@ae.mpg.de} \\
  \And 
  Ofer Tchernichovski \\
  Hunter College CUNY \\
  The CUNY Graduate Center \\
  \texttt{otcherni@hunter.cuny.edu} \\
  \And 
  Pauline Larrouy-Maestri \\
  Max Planck Institute for Empirical Aesthetics \\
  Frankfurt \\
  \texttt{pauline.larrouy-maestri@ae.mpg.de} \\
  \And 
  Nori Jacoby \\
  Max Planck Institute for Empirical Aesthetics \\
  Frankfurt \\
  \texttt{nori.jacoby@ae.mpg.de} \\
}

\defaultbibliographystyle{ieeetr}
\defaultbibliography{references}

\begin{document}
\begin{bibunit}

\maketitle

\begin{abstract}
 A core problem in cognitive science and machine learning is to understand how humans derive semantic representations from perceptual objects, such as color from an apple, pleasantness from a musical chord, or seriousness from a face. Markov Chain Monte Carlo with People (MCMCP) is a prominent method for studying such representations, in which participants are presented with binary choice trials constructed such that the decisions follow a Markov Chain Monte Carlo acceptance rule. However, while MCMCP has strong asymptotic properties, its binary choice paradigm generates relatively little information per trial, and its local proposal function makes it slow to explore the parameter space and find the modes of the distribution. Here we therefore generalize MCMCP to a continuous-sampling paradigm, where in each iteration the participant uses a slider to continuously manipulate a single stimulus dimension to optimize a given criterion such as ‘pleasantness’. We formulate both methods from a utility-theory perspective, and show that the new method can be interpreted as ‘Gibbs Sampling with People’ (GSP). Further, we introduce an aggregation parameter to the transition step, and show that this parameter can be manipulated to flexibly shift between Gibbs sampling and deterministic optimization. In an initial study, we show GSP clearly outperforming MCMCP; we then show that GSP provides novel and interpretable results in three other domains, namely musical chords, vocal emotions, and faces. We validate these results through large-scale perceptual rating experiments. The final experiments use GSP to navigate the latent space of a state-of-the-art image synthesis network (StyleGAN), a promising approach for applying GSP to high-dimensional perceptual spaces.
 We conclude by discussing future cognitive applications and ethical implications.
\end{abstract}

\section{Introduction}

Humans continuously extract semantic representations 
from complex perceptual inputs, re-expressing them as meaningful information that can be efficiently communicated to primary cognitive processes such as memory, decision-making, and language \cite{Anderson1990, Rosch1975, Jones2015}. Effective semantic representation seems to be a prerequisite for intelligent behavior, and is correspondingly a core topic of study in both cognitive science and machine learning \cite{Griffiths2007Topics, mesnil2014learning, wang2017alternative}.

One way of studying semantic representations in humans is to present participants with stimuli that exhaustively sample from a stimulus space (e.g., the space of visible colors) and ask them 
to evaluate these stimuli
(e.g., \citep{kay2009world}). 
Unfortunately, this method works poorly for high-dimensional stimuli whose parameter spaces are too large to explore exhaustively. An alternative approach is to hand-construct stimulus sets to test specific hypotheses about semantic representations (e.g., that slow melodies tend to sound sad, \citep{juslin2010musical}); 
however, this approach relies heavily on prior domain knowledge, and is poorly suited to exploratory research.

Markov Chain Monte Carlo with People (MCMCP) addresses this problem \citep{sanborn2008markov, sanborn2010uncovering, sanborn2015exploring}. 
MCMCP takes as input a stimulus space (e.g., visible colors) 
and a target semantic category (e.g., `danger’). 
In each trial, participants are presented with two stimuli
and are asked which comes from the category.
By virtue of MCMCP's adaptive procedure, 
stimulus selection becomes progressively biased
towards parts of the stimulus space that represent the category.
The resulting process iteratively characterizes the 
subjective mapping
between the stimulus space and the semantic concept
for a given participant or participant group.
The technique provides a way for cognitive scientists to systematically quantify subjective aspects of perception, 
for example the way in which participants from a 
particular musical culture hear certain chords as `consonant’, 
or the way in which participants have certain subjective ideas 
of what a ‘serious’ face looks like.
The approach has been shown to outperform reverse correlation, a related non-adaptive method for mapping semantic categories to perceptual spaces \citep{martin2012testing}.

According to the underlying theory, MCMCP converges asymptotically to the participant's internal probabilistic representation of a given semantic category within a stimulus space \citep{sanborn2008markov}. However, its practical convergence rate is limited for several reasons. The first concerns the response interface: MCMCP is traditionally limited to binary choice responses, which can only provide a small amount of information per trial (1 bit), much less than the theoretical limit of other response interfaces (e.g., sliders). The second depends on the proposal function that generates successive stimuli: a too-narrow proposal function makes the process slow to find the modes of the distribution, whereas a too-wide proposal function makes it harder for the process to estimate the mode with much precision \citep{gilks1996,hsu2012identifying}. 

Here we present a new technique for addressing these problems, termed Gibbs Sampling with People (GSP). While MCMCP corresponds to a human instantiation of the Metropolis-Hastings MCMC sampler, GSP corresponds to a human instantiation of a Gibbs sampler. 
Crucially, unlike \citep{griffiths2018subjective}, GSP has participants respond with a continuous slider rather than a binary choice. This has two effects: first, it substantially increases the upper bound of information per trial, and second, it eliminates the need to calibrate a proposal function. 
We further show how GSP can be formulated in utility theory, thereby generalizing the approach from discrete to continuous semantic representations, and we show how GSP can be shifted towards deterministic optimization through an aggregation process.




This paper continues with a review of MCMCP and a theoretical account of GSP.
We then describe four studies applying GSP to various visual and auditory domains, ranging from simple low-dimensional problems to complex high-dimensional problems parameterized by deep neural networks. 
These studies include experiments directly implementing GSP and MCMCP, control experiments investigating different hyperparameters, and validation experiments for the generated outputs. All combined, these 25 experiments represent data from 5,178 human participants.\footnote{Appendices, code, and raw data are hosted at \url{https://doi.org/10.17605/OSF.IO/RZK4S}.}

\section{Theory}
\subsection{MCMC and MCMCP}

MCMC is a procedure for sampling from distributions whose normalization constants are impractical to compute directly. It works by constructing a Markov chain whose stationary distribution corresponds to the probability distribution of interest; given enough samples from this Markov chain, it is then possible to approximate the probability distribution arbitrarily closely. 

The MCMC algorithm may be performed by choosing an arbitrary initial Markov state $x$ from the parameter space, then repeating the following steps until convergence:
(1) Sample a candidate $x^\ast$ for the next state of the Markov chain according to some \textit{proposal function} $q(x^\ast;x)$;
(2) Decide whether to accept this candidate according to an appropriate \textit{acceptance function} $A(x^\ast;x)$ constructed to reflect the probability distribution of interest.
In the case of a symmetric proposal function, the acceptance function takes a simple form known as the Barker acceptance function and is given by $A(x^\ast;x)=\pi(x^\ast)/(\pi(x)+\pi(x^\ast))$ where $\pi(x)$ is the target distribution \citep{barker1965monte}.

In MCMCP the acceptance function is replaced with a human participant, whose task is to choose between the current state and the candidate state \citep{sanborn2008markov}. The trick is then to frame this task such that the participant's choices correspond to the acceptance function for an interesting probability distribution. 
The solution presented in the original MCMCP paper is to tell the participant that one of the stimuli comes from a class distribution (e.g., cats), and one of them comes from an unknown category. The authors suppose that the participant computes the posterior probability of class membership assuming a locally uniform likelihood for the alternative class, and then selects a stimulus with probability proportional to its posterior probability of class membership. Under these assumptions, the participant's behaviour can be shown to correspond to the classic Barker acceptance function where the underlying probability distribution is the likelihood function for the class being judged.

This formulation is elegant but it has two important limitations. First, it can only be applied to semantic representations that take categorical forms; the derivation does not make sense for continuous semantic representations (e.g., pleasantness). Second, it assumes that participants make their choices with probabilities equal to their posterior probabilities of class membership (a process termed \textit{probability matching}) as opposed to the Bayes-optimal strategy of deterministically maximizing this posterior probability \citep{vulkan2000economist}. Humans do indeed seem to exhibit probability matching in certain contexts, but a convincing cognitive model ought to explain how this sub-optimal process arises \citep{shanks2002re}.

Here we reformulate MCMCP (and later GSP) without these limitations. We suppose that the participant is asked to select the stimulus that best matches a given criterion \(C\); for example `select the most pleasant chord' or `select the color that most resembles lavender'. We suppose that the participant performs this task by extracting a \textit{utility} value for each stimulus, and selecting the stimulus with the maximum utility \citep{McFadden1974}. In the case of the class membership tasks typically used in MCMCP, we might hypothesize that the utility corresponds to the subjective likelihood of the stimulus conditioned on the class of interest; however, in the general case the utility function need not necessarily correspond to a probability distribution. The utility value is however assumed to have a deterministic component and a noise component, namely $U_{i}=\ell_{i}+n_{i}$, where $\ell_i$ is the deterministic utility of stimulus $i$, and $n_{i}$ is the associated noise. This noise component can capture participant-level noise from sensory \citep{weiss2002motion, wei2015bayesian} and cognitive \citep{wei2015bayesian, Sanborn2006} processes, as well as population-level noise corresponding to individual differences in the utility function \citep{McFadden1974}. In the case where the noise components are i.i.d.\ and have an extreme value distribution common to discrete choice models, it can be shown that the probability of selecting a given stimulus $s_1$ is equal to
$(1+\exp({-\gamma(\ell_1-\ell_2)}))^{-1}$
where $\gamma^{-1}$  corresponds to the scale parameter of the noise component. If the utility is assigned based on subjective likelihood $\ell_i=\log p(s_i|C)$, then this equation 
reproduces the Barker acceptance function with target distribution $\pi(s) \propto p^\gamma(s|C)$. This justifies MCMCP for an optimal observer with a noisy utility function (for proof and discussion see Appendix \ref{appendix-mcmcp}).

\subsection{Gibbs Sampling with People}

Gibbs sampling is an alternative approach for sampling from probability distributions \citep{gelfand1990sampling}, defined as follows. Let $p(z_1,\dots,z_n)$ be a target distribution over an $n$-dimensional state space from which one would like to sample, 
and choose a starting vector state $\mathbf{z}^{(1)} = (z_1^{(1)}, \ldots, z_n^{(1)})$.
Then, circularly update coordinates by sampling from $p(z^{(i+1)}_{k}|z_1^{(i+1)},\dots,z^{(i+1)}_{k-1},z^{(i)}_{k+1},\dots,z^{(i)}_{n})$.

In GSP the participant provides the coordinate updates. This is achieved by presenting the participant with a slider that is associated with the current stimulus dimension $z_k$ and instructing the participant to move the slider to maximize a certain criterion, such as the pleasantness of a sound or the resemblance of a fruit to a strawberry. To analyze the decision step, let us discretize the slider into a set of points $\{{z_k^{i}}\}_i$  and let $z_{-k}$ denote the other fixed dimensions. As before, suppose that each point on the slider is associated with a utility that contains both a deterministic $\ell(z_k^i,z_{-k})$ and a noise $n_i$ component, and suppose that the participant chooses the slider location that maximizes the utility. Then, under similar assumptions to those made in the MCMCP case, the probability distribution over slider locations is 
\begin{equation}
p(\mbox{choose }i)
=p(z_k^i|z_{-k})
=\frac{e^{\gamma\ell(z_k^i,z_{-k})}}{\sum_{j}e^{\gamma\ell(z_k^j,z_{-k})}}
\label{eq:GSPstep}
\end{equation}
and as the granularity of the slider tends to infinity, the denominator becomes a marginal, and GSP becomes a sampler from $p(\textbf{z}) \propto e^{\gamma\ell(\textbf{z})}$ (for proof and discussion see Appendix \ref{appendix-gsp}).


As with MCMCP, GSP can be used to explore either categorical or continuous semantic representations. In the former case, the experimenter might ask a question like `adjust the slider until the image most resembles a dog', and the participant's utility function might correspond to the log probability of the image given the class, $\ell(\textbf{z})=\log p(\mathbf{z} \vert C)$; in this case the sampler's stationary distribution will be proportional to $p^\gamma (\mathbf{z} \vert C)$. In continuous semantic representations, the utility function may not correspond to a probability distribution, and the interpretation would simply be that the sampler explores different regions of the space in proportion to their exponentiated utility $ e^{\gamma\ell(\textbf{z})}$.

The noise parameter $\gamma ^ {-1}$ is important for the behavior of the sampler. As $\gamma ^ {-1} \rightarrow 0$ , the choice distribution becomes increasingly peaked around the highest utility item on the slider, shifting thus the sampler into an optimization regime (specifically, coordinate descent). Typically we are interested in minimizing $\gamma ^ {-1}$, so as to maximize the utility of the samples (mode seeking); however, some noise is still desirable because it helps drive exploration of the utility space.


There are two main ways to reduce the effective noise, $\gamma ^ {-1}$.
One approach is to estimate the joint distribution $p(\textbf{z}) \propto e^{\gamma\ell(\textbf{z})} $ by fitting a kernel density estimate (KDE) to the GSP samples, then simulating $\gamma ^ {-1} \rightarrow 0$ by taking the distribution's mode.
For simple distributions, this mode can also be estimated by averaging over samples.
However, neither KDEs nor averaging are well-suited to complex high-dimensional spaces, where the joint distribution is hard to estimate reliably.

An alternative approach is to manipulate the Gibbs sampler itself. Specifically, suppose that we collect multiple samples from the conditional distribution of a given step of the Gibbs sampler
$p(\mbox{choose } i) \propto e^{\gamma\ell(z_k^i,z_{-k})}$, 
and then simulate $\gamma ^ {-1} \rightarrow 0$ by returning the peak of the one-dimensional KDE from these samples (or potentially the sample mean). This will in turn simulate $\gamma ^ {-1} \rightarrow 0$ for the joint distribution $p(\textbf{z}) \propto e^{\gamma\ell(\textbf{z})}$ as produced by the Gibbs sampler. The practical advantages of this approach are that (a) we restrict density estimation to a more tractable one-dimensional case, and (b) the same stimulus can be re-used for multiple trials, which can be useful when stimuli are slow to create. We explore both KDE and mean aggregation in this paper.

There are several ways to assign the iterations of a GSP or MCMCP chain to human participants. In \textit{within-participant} chains, the entire chain is completed by just one participant, and the resulting samples reflect the semantic representations of that single participant. In \textit{across-participant} chains, each iteration comes from a different participant, and the samples then reflect shared semantic representations across participants (Fig.\ \ref{fig:chain-types}). 
While within-participant chains can theoretically be used 
to study individual differences,
here we focus on studying representations at the level of the
participant group, using both chain types
and averaging over participants where appropriate.

Researchers interpreting MCMCP and GSP results must think carefully both about the definition of the stimulus space and of the participant group. For example, if the stimulus space only includes male voices, the results may not be generalizable to female voices. Similarly, if the participant group comprises solely US participants, then the results may not be generalizable to Japanese participants. Of course, these issues are by no means limited to MCMCP and GSP, but apply rather to the majority of psychological research. We will revisit these matters below.

A related paradigm with a Gibbs sampler interpretation is \textit{serial reproduction}, where one participant's imitation of a stimulus becomes the next stimulus in a transmission chain \citep{jacoby2017integer, kempe2015structure, Xu2010,  kirby2008cumulative, verhoef2014emergence, edmiston2018repeated, braun2006evidence}. 
However, serial reproduction is limited to percepts that can be entirely reproduced in a single trial (e.g., spoken sentences, \citep{braun2006evidence}). 
In contrast, GSP participants only ever have to manipulate one stimulus dimension at a time, even if the stimulus itself is high-dimensional. This allows GSP to explore much richer stimulus spaces.
A second related paradigm with a Gibbs sampler interpretation is described by \citep{griffiths2018subjective}, studying subjective randomness by having participants impute missing parts of coin-flip sequences.
Our approach differs in soliciting continuous rather than discrete judgments.
A third related paradigm is the multidimensional method of adjustment, where participants simultaneously adjust multiple sliders to make a stimulus match a certain criterion (e.g., \citep{grimaud2019emotecontrol}). GSP differs from the latter in providing a principled way to share the task between participants, and a coherent probabilistic model relating slider movements to the utility function.

\section{Studies}

\subsection{Color} \label{section:color}

Our first study concerns a particularly low-dimensional perceptual space: color. This kind of perceptual space should provide a useful sanity check for any semantic sampling procedure: if a procedure fails here, it is surely even less likely to succeed in high-dimensional perceptual spaces.\footnote{Additional methods, results, and demographic information for all experiments are provided in the Appendices.}

We tested our sampling methods on recovering the colors associated with eight words: `chocolate', `cloud', `eggshell', `grass', `lavender', `lemon', `strawberry', and `sunset'. We parameterized color space using the perceptually oriented HSL scheme \citep{joblove1978color}, where each color is encoded as three integers: hue, saturation, and lightness, taking values in [0, 360], [0, 100], and [0, 100] respectively.

\begin{figure}
  \centering
  \includegraphics[width=1.0\linewidth]{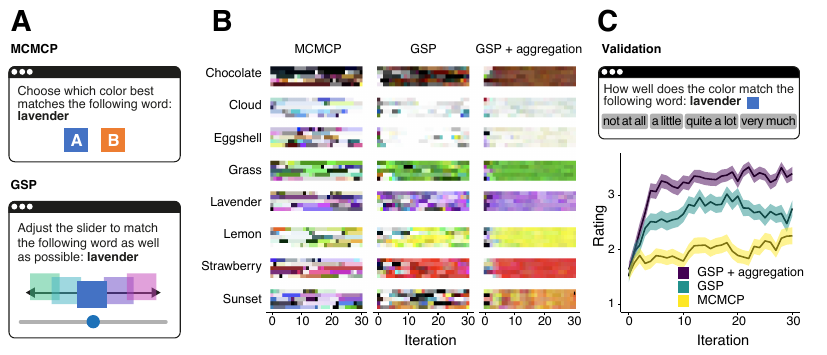}
  \caption{Sampling color representations. \textbf{A}: MCMCP/GSP instructions. \textbf{B}: Generated samples. \hbox{\textbf{C}: Task} and results for the validation experiment (95\% confidence intervals over participants).}
  \label{fig:color}
\end{figure}

The first sampling method was MCMCP, implemented with a Gaussian proposal function of standard deviation 30. The second method was standard GSP. The third method was aggregated GSP, collecting 10 slider responses for each iteration, and propagating the mean response to the next iteration.
Each method was evaluated using across-participant chains of length 30, with five chains per color category,
with each chain's starting location sampled from a uniform 
distribution over the color space
(Exp.\ 1a, 1b, 1c). All participants (\textit{N} = 422) were recruited from Amazon Mechanical Turk (AMT) and pre-screened with a color-blindness test and a color-vocabulary test before continuing with the online experiment (Appendix \ref{appendix-color}). Each participant contributed up to 40 trials for a given method. In each case, the participant was presented with a word (e.g., `lavender'), and asked to choose the color that best matched that word with either a binary choice interface (MCMCP) or a slider (GSP) (Fig.\ \ref{fig:color}A).

There are several ways that one could evaluate the success of an MCMCP or GSP procedure.
Here we follow previous work 
by having participants rate how well samples match the target category \citep{martin2012testing},
but see Appendix \ref{appendix-color} for an alternative analysis.
We elicited c.\ 5.2 ratings per sample from a new participant group (\textit{N} = 322, Exp.\ 1d); participants were presented with the target word from the original chain, and asked to judge how well the color matched this word on a scale from 1 (not at all) to 4 (very much). 
The results indicate a clear advantage for GSP over MCMCP, with GSP converging faster and on higher ratings; they also show that aggregation robustly improved ratings (Fig.\ \ref{fig:color}B, \ref{fig:color}C). 
Inspecting Fig.\ \ref{fig:color}B, it is clear that many MCMCP samples poorly reflected their semantic category; meanwhile, GSP produced considerably fewer poor samples, and aggregated GSP even fewer. 
Investigating further, we found that the poor performance of MCMCP persisted when
(a) normalizing for the longer duration of GSP trials (Fig.\ \ref{fig:color-validation-profile-normed}),
(b) trying different proposal widths
(Exp.\ 1e, 153 participants, Fig.\ \ref{fig:mcmcp-explore}), 
(c) using different questions
(Exp.\ 1f, 190 participants, Fig.\ \ref{fig:color-questions}),
(d) implementing within-experiment aggregation
(Exp.\ 1g, 1h, 572 participants),
(e) implementing post-hoc aggregation (Fig.\ \ref{fig:cumulative-color}),
and (f) accounting for the trade-off between mode-seeking
and exploration
(Exp.\ 1i, 270 participants, Fig.\ \ref{fig:color-1d-kernels}).
The implication is that, when the stimulus space is well-parameterized, GSP substantially improves sampling quality over MCMCP.
In addition, it is clear that aggregation improves sampling quality still more at the cost of additional participant trials.

As an exercise, it is useful to reflect on how the stimulus space and the participants might constrain the generalizability of these results. The stimulus space presents little problem; every visible color has a close neighbor in the HSL scheme used here. However, the results should not be expected to generalize globally, given well-documented cross-cultural variations in color-naming \citep{kay2009world}.


\subsection{Emotional prosody} \label{section:prosody}

This study concerns a long-standing psychological question: how the way that a sentence is spoken (its \textit{prosody}) communicates the speaker's emotional state \citep{banse1996}. Prior research mostly depends on recordings of actors expressing particular emotions, but such stereotypical recordings might not fully reflect natural emotion perception \citep{banziger2015}. GSP provides a way to study prosody perception without actors, instead generating emotional prosody directly from the perceptual judgments of listeners.

\begin{figure}
  \centering
  \includegraphics[width=1.0\linewidth]{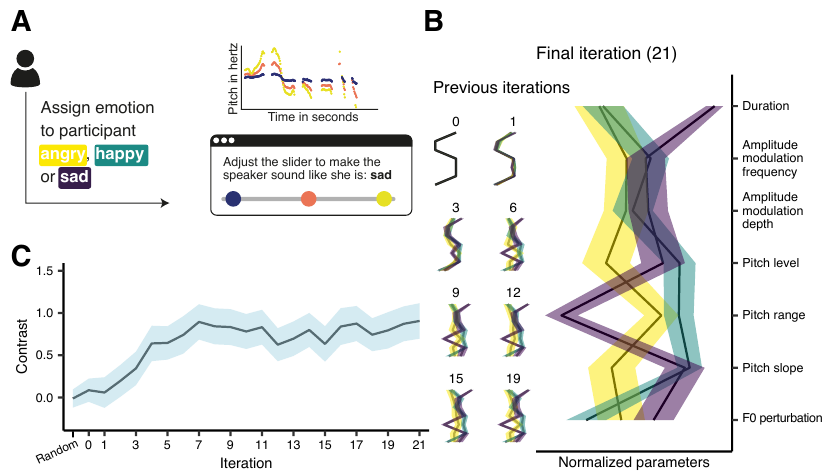}
  \caption{Sampling emotional prosody. \textbf{A}: Overview of the GSP task. \textbf{B}: Mean feature values by iteration. \textbf{C}: Mean `contrast' ratings, corresponding to the mean rating for the `correct' emotion minus the mean rating for the `incorrect' emotions (95\% confidence intervals over participants).}
  \label{fig:emotion}
\end{figure}

We began with three sentences from the Harvard sentence corpus \citep{ieee} recorded by a female speaker \citep{demonte2019}, chosen to facilitate comparison with previous research; these sentences are phonologically balanced and semantically neutral. We defined our stimulus space in terms of seven parametric manipulations, corresponding to \textit{duration} (speeding up or slowing down the fragment), \textit{intensity variation} (rate and depth) and \textit{pitch} (absolute level, range, slope, and F0 perturbation). We explored this space using 220 within-participant GSP chains, 
each comprising 21 iterations,
and each beginning with the original unaltered recording
(Exp.\ 2a). 
Participants (\textit{N} = 110) were recruited from AMT, pre-screened with the audio test of \cite{woods2017headphone}, and each randomly assigned to either `anger', `happiness', or `sadness' (Fig.\ \ref{fig:emotion}A). Each participant contributed two chains corresponding to different sentences.

Fig.\ \ref{fig:emotion}B plots mean feature values for the different emotional categories. Sad speech was marked by long duration, reduced pitch range, shallow pitch slope, and high F0 perturbation. Happiness had short duration, increased mean pitch, shallow pitch slope, and high pitch range. Anger had short duration, low mean pitch, falling pitch slope, and high pitch range. These characterizations are generally consistent with previous research (e.g., \citep{laukka2016}). We also observed interesting patterns of feature correlations. For example, we found duration and F0 perturbation to be correlated for sadness (\textit{r} = .28) but not for the other emotions (anger: \textit{r} = $-$.03, happiness: \textit{r} = .00); in contrast we found that pitch level and pitch slope were positively correlated in all three emotions (Fig.\ \ref{fig:emo-cor}). This suggests a new way to explore the perceptual spaces of perceived emotions, contrasting with previous literature that mainly focuses on unique contributions of single dimensions.

We then evaluated the resulting samples with a new participant group (\textit{N} = 161), who rated how well samples matched each emotion on a four-point scale, producing c.\ 5.4 ratings per stimulus (Exp.\ 2c). Ratings increased steadily for the first sweep of the parameter vector and then plateaued with a reliable mean contrast of 0.9 points (Fig.\ \ref{fig:emotion}C).
We also replicated the results with across- instead of within-participant chains (Exp.\ 2b, 2d, 210 participants, Fig.\ \ref{fig:emo-sup}A).

These results imply that GSP is effective for exploring emotional prosody, and for generating emotional stimuli without the confounds of acted recordings.
Nonetheless, there are clear ways in which this work could be extended.
The stimulus space was defined by a limited set of manipulations, such as mean pitch, pitch slope, and F0 perturbation; this set could be extended to include for example spectral features or more granular pitch manipulations \citep{juslin2003communication,scherer2019acoustic}. The stimuli all correspond to English sentences, and the participants were all US participants; our results should not be assumed to generalize outside this cultural context \citep{laukka2020cross,paulmann2014cross}. Moreover, all stimuli were synthesized with a female voice, so the results should not be assumed to generalize to male speakers.

\subsection{Musical chords} \label{section:chords}

Our third study concerns the subjective pleasantness of musical pitch combinations, or \textit{chords}. For Westerners, this domain is highly multimodal, containing many prototypes of `pleasant' (or `consonant') chords. Exhaustively exploring this continuous space is difficult for conventional methods, and so far such research has been limited to single pitch intervals or to specific tuning systems \cite{HarrisonConsonance, lahdelma2020cultural, Plomp1965, Sethares2005}. Here we investigate whether GSP can help us to characterize the continuous space of \textit{pairs} of pitch intervals without restricting stimuli to a given tuning system.

\begin{figure}
  \centering
  \includegraphics[width=1.0\linewidth]{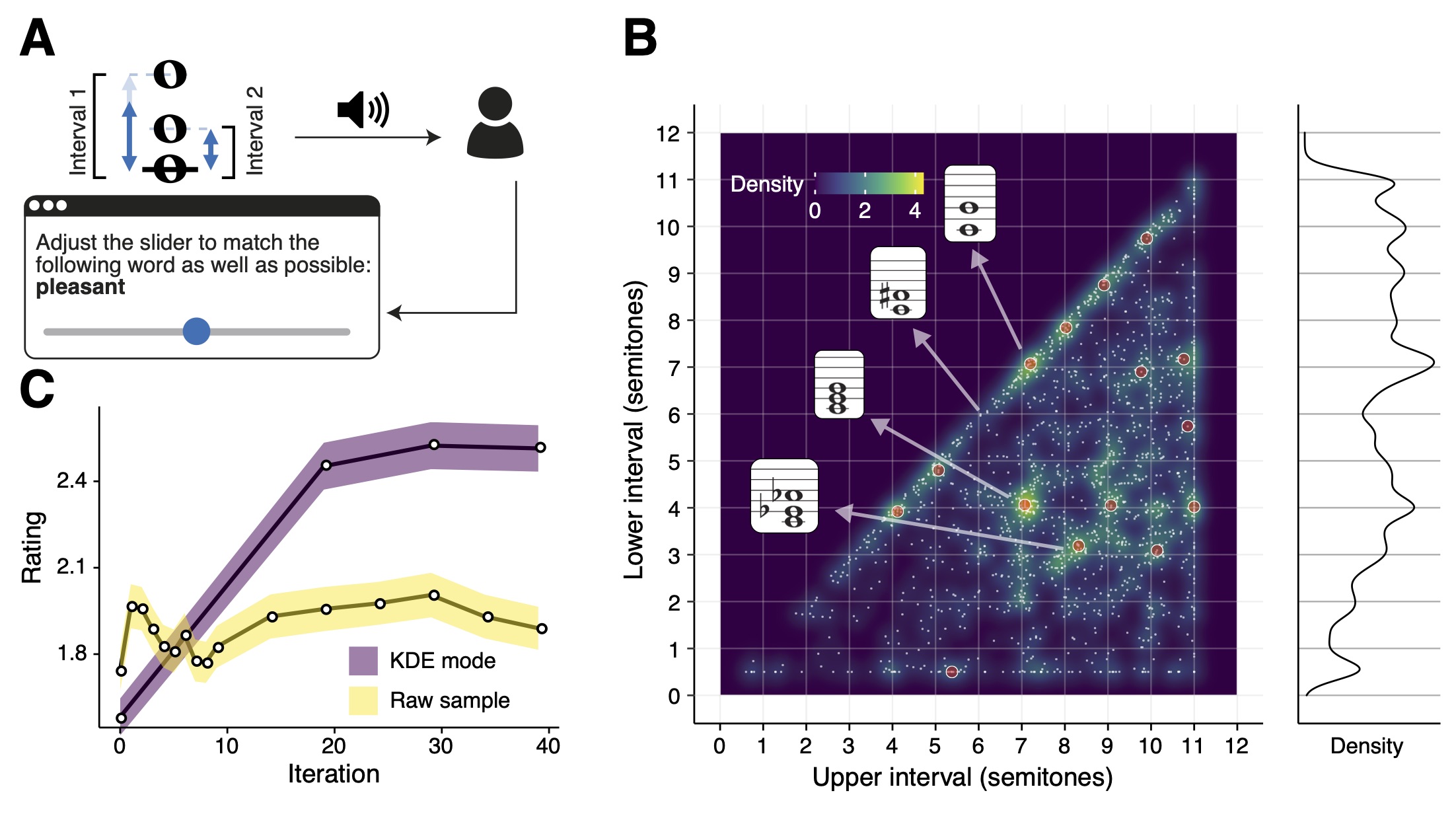}
  \caption{GSP over musical chords. \textbf{A}: Schematic illustration of the experimental task.
  \textbf{B}: KDE over iterations 10 to 39, with density expressed relative to a uniform distribution, the top 15 modes marked by red dots, all plotted alongside the marginal distribution of lower and upper intervals combined.
  \textbf{C}: Validation ratings by iteration (95\% confidence intervals over responses).}
  \label{fig:chord}
\end{figure}

Our stimulus space comprised two continuous intervals, specifying the logarithmic distance from the bass tone in the range 0.5--11 \citep{jacoby2019universal}. 
The standard Western tuning system corresponds to integer coordinates in this space.
We explored this space with 50 across-participant GSP chains
of length 40,
whose starting locations were sampled from a uniform distribution 
over the stimulus space.
The participants (\textit{N} = 134) were recruited from 
AMT and pre-screened with the audio test of \citep{woods2017headphone} (Exp.\ 3a). These participants were instructed to make each chord as `pleasant' as possible (Fig.\ \ref{fig:chord}A). 
In a subsequent validation experiment, participants (\textit{N} = 168) rated pleasantness for samples from (a) the empirical distribution and (b) the top five modes of KDEs applied to raw samples from iteration 10 onwards
(Exp.\ 3b). Each condition received 662 ratings with up to 80 ratings per participant.

Ratings increased clearly as a function of iteration, with KDE modes scoring significantly higher than raw samples.
The KDEs display a rich structure that replicates and extends prior research (Fig.\ \ref{fig:chord}B) \cite{HarrisonConsonance, lahdelma2020cultural, Plomp1965, Sethares2005}. In particular, the 1D KDE shows clear integer peaks corresponding to the Western tuning system, with dips at the semitone (1) and tritone (6); the 2D KDE additionally shows peaks at various prototypical sonorities from Western music, such as the major triad (4, 7), the first inversion of the major triad (3, 8), and a dominant seventh chord with omitted third (7, 10) (see e.g., \citep{bowling2018vocal}; see also Fig.\ \ref{fig:dynamicMarginal}). 
These results imply that GSP is effective for exploring continuous musical spaces.

Our stimulus space only contained three-tone chords, but of course real music contains many different varieties of chords. Our chord tones were synthesized using artificial harmonic complex tones; though such tones are commonly used in prior research \citep{mcdermott2016indifference}, real music contains many different kinds of tones, some of which have different consonance profiles \citep{Sethares2005}. Moreover, our participant group comprised mostly US and Indian participants, yet consonance perception is known to vary cross-culturally \citep{mcdermott2016indifference}. Future work should explore how our results vary as a function of these variables.


\subsection{Faces} \label{section:faces}
Our final study addresses a particularly high-dimensional domain: images of human faces. Such images would be too high-dimensional for GSP to manipulate in their raw form, so we instead parameterize the stimuli with a generative model. State-of-the-art image synthesis models typically still have high-dimensional parameter spaces, but here we build on recent work showing that the latent space of these models can be effectively navigated using principal component analysis (PCA) \citep{ganspace}. Following \citep{ganspace}, we apply this approach to the generative adversarial network `StyleGAN' \cite{karras2019style, stylegan2}, pretrained on the FFHQ dataset of faces from Flickr \cite{karras2019style}, and applying PCA to the intermediate latent code (termed $\mathbf{w}$ in the original papers). 
We used the top 10 PCA components to parameterize our stimulus space, allowing these components to vary up to two standard deviations from the mean, and fixing the input latent code ($\mathbf{z}$ in the original papers) to the mean to control variability.

\begin{figure}
  \centering
  \includegraphics[width=1.0\linewidth]{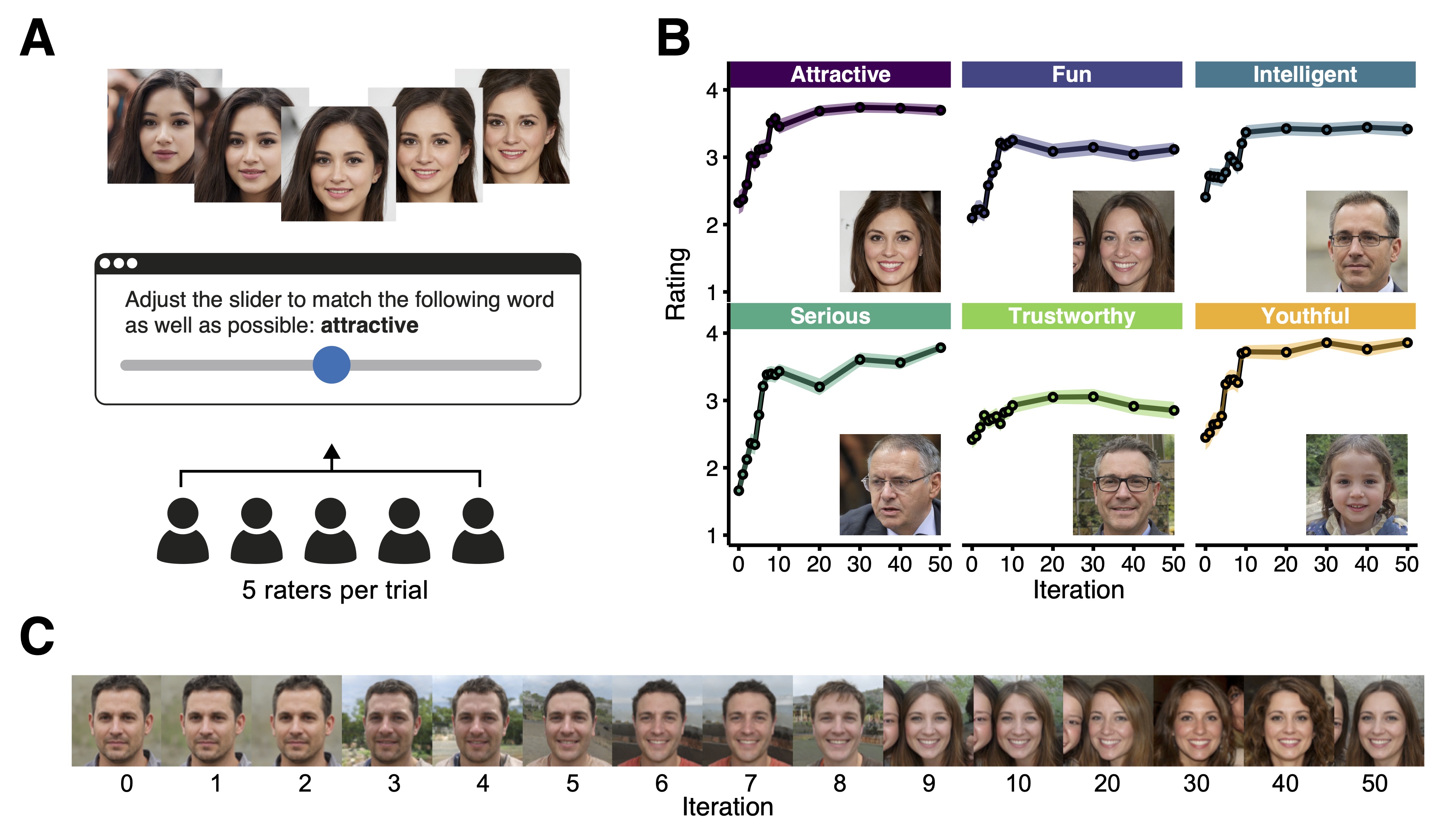}
  \caption{Sampling facial representations. \textbf{A}: Instructions for the GSP task. \textbf{B}: Results of the validation experiment, including final samples for each target adjective (95\% confidence intervals over participants). \textbf{C}: Example GSP chain for `fun', with samples ordered by iteration.}
  \label{fig:face}
\end{figure}

We used the resulting generative model to explore subjective stereotypes for the following adjectives:
`attractive', `fun', `intelligent', `serious', `trustworthy', and `youthful',
with these choices informed by prior literature (e.g., \citep{brinkman2017visualising}).
We constructed 18 across-participant GSP chains of length 50
with uniformly sampled starting locations 
and three chains for each adjective (Fig.\ \ref{fig:face}A, Exp.\ 4a).
We used 293 US participants from AMT,
aggregating 5 trials per iteration using the arithmetic mean. We then evaluated the generated samples with a rating experiment, following the same procedure as the color experiment but collecting c.\ 52.1 ratings per sample from 179 US participants (Exp.\ 4b).

The results are illustrated in Fig.\ \ref{fig:face}B--C. The GSP chains converged on highly rated samples remarkably quickly, with one full sweep of the 10 dimensions being sufficient to effectively capture the target categories as evaluated by the validation experiment. 
This implies that GSP can indeed successfully navigate StyleGAN's generative space. 
Follow-up experiments found similar success with different dimensionality reduction techniques and aggregation methods (Exp.\ 4c--f, Appendix \ref{appendix-faces}).

Samples from the GSP process will inherit certain biases from the StyleGAN model. For example, if male faces are over-represented in StyleGAN samples, they are likely to be over-represented in the GSP samples; likewise, if StyleGAN samples contain predominantly young female faces and old male faces, then GSP samples for `youthful' are likely to be biased towards female faces. To examine such biases, we conducted a follow-up experiment analyzing the distribution of different features 
as subjectively rated by online participants (Exp.\ 4g, Appendix \ref{appendix-faces}).
The results indicate that StyleGAN's training dataset already contains significant biases that are
propagated through the modeling pipeline, 
and potentially contribute to the prevalence of white faces in the GSP samples, as well as gender associations for the different targets.
These findings indicate the importance of interpreting GSP results in the context of their associated generative models, and of sourcing less biased training datasets for future cognitive applications \citep{scheuerman2020we}; 
though StyleGAN's FFHQ dataset may be more diverse than many competing machine-learning datasets, it is clearly not bias-free. 
Our results will also reflect the stereotypes held by our participant group;
repeating this method with different participant groups could yield interesting hypotheses concerning how facial stereotypes vary across different demographics and cultures.
Appendix \ref{appendix-faces} describes initial experiments in this line with participant groups differentiated by gender and location (Exp.\ 4d, 4i, 4j).

\section{Summary and conclusion}

We have presented GSP, a new technique for extracting semantic representations from human participants. GSP organizes these participants into virtual Gibbs samplers, and thereby generates stimuli from the perceptual space associated with a given semantic representation.
We have shown how this technique can recover semantic representations for a variety of perceptual domains, including color, emotional prosody, musical chords, and faces. The richness of the derived representations is compelling, and suggests many future applications in cognitive and social sciences.

GSP has several features that seem to help it converge quickly on high-quality samples. One is its continuous-slider interface, which can deliver much more information per trial than the binary choice method used by MCMCP. A second is its lack of tuning parameter, which reduces the resources required to develop a workable experiment. A third is the way in which it manipulates a single stimulus dimension at a time: it is plausible that participants find it easier to evaluate differences between stimuli when the stimuli differ on just a single perceptual dimension.

By formulating GSP and MCMCP in utility theory \cite{McFadden1974}
we enable both methods to be applied to continuous as well as categorical semantic representations, while relaxing assumptions about the participant's prior and response noise. By incorporating aggregation into the conditional part of the Gibbs sampler, we increase the participant-to-stimulus ratio and thereby make GSP practical for stimuli that take a long time to generate, with the useful byproduct of averaging out perceptual noise.


The final study showed how GSP can be used to navigate the latent space of deep neural synthesis models.
The important prerequisite is finding a relatively low-dimensional basis for the network for GSP to parameterize; fortunately, it seems that relatively simple techniques such as PCA can sometimes suffice for this task \citep{ganspace}. 
This approach has clear potential for helping cognitive scientists to study semantic representations in high-dimensional perceptual spaces.




\section*{Broader Impact}



This research extends the methods available to cognitive scientists who seek to characterize semantic representations in human participants.  In particular, the proposed method facilitates studying much richer perceptual spaces (both in terms of dimensionality and in terms of granularity) than can be explored effectively with conventional methods. 

Our research group is particularly interested in using GSP to study cross-cultural differences in perception \citep{jacoby2019universal,jacoby2020cross,jacoby2017integer}. In this context, exploratory techniques such as GSP are particularly useful, because they can generate valuable cognitive insights without specifying a constrained hypothesis space \textit{a priori}. Previous work using slider interfaces with cross-cultural populations makes us relatively confident that GSP could be applied cross-culturally \citep{sievers2013music}, as long as sufficient care is taken to ensure that the task is understood properly by the participants.
Addressing cross-cultural populations in this way can help to ameliorate cognitive science's longstanding bias towards participants from WEIRD (Western, Educated, Industrial, Rich and Democratic) backgrounds \citep{henrich2010most}.



It is important to identify potential pitfalls in applying GSP, especially when such activities have adverse ethical implications. We give three recommendations below for avoiding such mistakes.

\textbf{Do not conflate subjective judgments with objective truth.} GSP is a tool for understanding participants’ subjective notions of particular semantic concepts. It does not necessarily reveal any objective truth about these concepts. This is particularly relevant in examples like our face study, where GSP is used to characterize perceived intelligence and trustworthiness. For example, GSP may suggest that participants associate glasses-wearing with intelligence: this does not mean that wearing glasses makes someone intelligent, or even that glasses wearing is necessarily associated with intelligence in the real world. Mistakes of this kind have the potential to perpetuate or amplify dangerous stereotypes in society, especially when the inferences concern race/ethnicity and gender;
such an approach has a regrettable history in the now-discredited field of physiognomy. 
Researchers using our method and related psychological methods should be aware of this negative history, and hold their own work to a higher ethical standard to avoid causing similar harm.
Consequently, GSP should not be used as tool for generating training datasets for machine-learning algorithms, or for fine-tuning the parameters or hyperparameters of such algorithms, unless the researcher makes it absolutely clear that the algorithm is being used to study human stereotypes rather than objective truths.

\textbf{Analyze, report, and ideally avoid potential biases.} Cognitive scientists must always be sensitive to potential biases in designing their stimuli and recruiting their participants. GSP is no exception to this principle. 
Our studies include examples of relatively simple and unbiased stimulus spaces (HSL colors; musical triads) as well as examples of relatively complex but potentially biased stimulus spaces (recordings of spoken sentences; images generated by the StyleGAN model). 
For practical reasons, our studies all used participants recruited from AMT; while this platform provides a relatively diverse participant group compared to the common practice of recruiting psychology students, it clearly does not represent the full diversity of the global population \citep{difallah2018demographics}, and our results are likely to reflect culturally dependent stereotypes as a result (e.g., the Western preference for musical chords with high harmonicity, \citep{mcdermott2016indifference}).
It is imperative that cognitive scientists remain vigilant concerning the potential harms of using non-diverse participant groups, both as regards making incorrect scientific conclusions and as regards perpetuating the under-representation of already marginalized parts of society \citep{henrich2010most}.
We discuss these issues on a case-by-case basis above, but future cognitive work using these methods should examine these issues in greater detail. For example, we did not gather detailed personal information about our participants on variables such as race/ethnicity due to privacy reasons, but it is important that future work studying facial stereotypes takes such variables into account. 

\textbf{Validate findings with rigorous hypothesis-driven experiments.} The power of combining GSP with deep generative models (e.g., StyleGAN) is that it enables the researcher to ask exploratory questions about complex naturalistic stimuli, such as `what do people think a serious face looks like?' However, the downside of this approach is that the technique is susceptible to inheriting hidden biases from the generative model \citep{grover2019fair}. 
It is therefore essential that cognitive research combining GSP with deep generative models should treat the results as exploratory, and ideally validate the results with well-controlled experiments that do not rely on the generative model. 

\begin{ack}
The authors are grateful to David Poeppel for general help and support, and to Alec Mitchell, Jesse Snyder, Jordan Suchow, Matthew Wilkes, and Sally Kleinfeldt for their support of the Dallinger project. We would also like to thank Roya Pakzad for advising us on ethical aspects of the project.
\end{ack}

\medskip

\small
\putbib
\end{bibunit}

\clearpage

\begin{appendices}

\begin{bibunit}

\section{Mathematical framework}\label{appendix-maths}

\setcounter{figure}{0}
\makeatletter 
\renewcommand{\thefigure}{S\@arabic\c@figure}
\makeatother

\setcounter{table}{0}
\makeatletter 
\renewcommand{\thetable}{S\@arabic\c@table}
\makeatother

In this section we present derivations of the formulas derived in the theoretical exposition of the main text. We start from a derivation of the acceptance function of MCMCP based on utility theory which we then generalize to GSP.

\subsection{MCMCP}\label{appendix-mcmcp}

To analyze the decision step of MCMCP, let us imagine that a participant is presented with two alternatives, $s_{1}$  and $s_{2}$, from which they are asked to choose according to some criterion $c$. In the context of utility theory, we suppose that the participant performs this task by extracting a \textit{utility} value for each stimulus, and selecting the stimulus with the maximum utility.  The utility value has two components, a deterministic component and a noise component, namely $U_{i}=\ell_{i}+n_{i}$, where $\ell_i$ is the deterministic utility of stimulus $i$, and $n_{i}$ is the associated noise. This noise component can capture intra-participant noise from sensory \citep{weiss2002motion, wei2015bayesian} and cognitive \citep{wei2015bayesian, Sanborn2006} processes, as well as inter-participant noise corresponding to individual differences in the utility function \citep{McFadden1974}. In the current derivation, we assume that the noises are i.i.d.\ and that they are Gumbel distributed (also known as type I extreme value), $n_{i}\sim \mbox{Gumbel}(\mu,\gamma^{-1})$. Gumbel distributions are commonly used in discrete choice models because they approximate Gaussian noise while possessing useful analytic properties \citep{train2009discrete}. From here, the probability of choosing, say $s_1$ , would be
\begin{equation}
p(\mbox{choose}\,s_1)=p(U_1>U_2)=p(n_2-n_1<\ell_1-\ell_2).
\label{eq:MCMCPstepexp}
\end{equation}
To proceed, we recall the following useful property of Gumbel distributions:  let $X_1\sim\mbox{Gumbel}(\mu_1,\gamma^{-1})$ and $X_2\sim\mbox{Gumbel}(\mu_2,\gamma^{-1})$ be two independent variables, then the difference is logistically distributed, namely, $X_1-X_2\sim\mbox{Logistic}(\mu_1-\mu_2,\gamma^{-1})$. Thus, we see that the right hand side of (\ref{eq:MCMCPstepexp}) is simply the cumulative distribution function of the logistic distribution $\mbox{Logistic}(0,\gamma^{-1})$, from which it follows that
\begin{equation}
p(\mbox{choose}\,s_1) = \frac{1}{1+e^{-\gamma(\ell_1 - \ell_2)}}
\end{equation}
which is the desired result (see Section 2.1 in the main paper). 

\subsection{GSP} \label{appendix-gsp}

Let us now generalize the analysis of MCMCP to GSP. Recall that in the GSP step, a participant is presented with a slider that is associated with an active dimension, say $z_k$, from which they are asked to select a value. To analyze the decision step, let us discretize the slider into a set of points $\{z_k^i\}_i$,  and let $z_{-k}$ denote the other fixed dimensions. Similar to the MCMCP case, we assume that the participant extracts a utility value for each stimulus along the slider, namely, $U_i=\ell(z_k^i,z_{-k})+n_i$, with the noise being i.i.d.\ and Gumbel distributed $n_i\sim\mbox{Gumbel}(\mu,\gamma^{-1})$, and we assume that they choose the alternative with the highest utility. Such a choice model is known in the literature as the multinomial logit \citep{train2009discrete}.  For completeness, let us derive the formula for the probability of choosing the alternative $z_k^i$ . We have

\begin{equation*}
\begin{split}
p(z_k^i|z_{-k})&=p\left(\bigcap_{j\ne i} U_i > U_j\right)\\
&=\int_{-\infty}^{\infty} d\epsilon p\left(\bigcap_{j\ne i} n_j  < \ell_i - \ell_j +n_i  \middle|n_i = \epsilon\right)p(n_i=\epsilon)\\
&=\int_{-\infty}^{\infty} d\epsilon \prod_{j\ne i}p(n_j < \ell_i - \ell_j + \epsilon)p(n_i = \epsilon)\\
&=\gamma\int_{-\infty}^{\infty} d\epsilon \exp\left\{-\left(1 + \sum_{j\ne i}e^{-\gamma(\ell_i - \ell_j)}\right)e^{-\gamma(\epsilon-\mu)}-\gamma(\epsilon-\mu)\right\}\\
&=\frac{e^{\gamma\ell_i}}{\sum_{j}e^{\gamma\ell_j}}
\end{split}
\end{equation*}

where in the third equality we used the fact that the noises are independent, and in the fourth equality we plugged in the standard formulas for the probability and cumulative distributions of the Gumbel distribution. The fifth equality follows from substituting $\epsilon^\prime=\gamma(\epsilon-\mu)$  and noticing that the sum over exponentiated utility differences is a positive number, so that the integral identity $\int_{-\infty}^{\infty}dx \exp\{-\lambda e^{-x}-x\}=1/\lambda$ holds. Thus, substituting $\ell_i = \ell(z_k^i,z_{-k})$ we arrive at the desired equation, that is, Equation (1) in the main paper. Notice also that in the case of two alternatives, this derivation recovers the acceptance function of MCMCP.

Both derivations of the MCMCP and GSP choice probabilities relied on two main assumptions regarding the nature of the noise: (a) it is i.i.d., and (b) it is Gumbel distributed. Starting from the latter, notice that the derivation of the GSP choice probability makes it clear how to generalize to other types of noise. Indeed, up to (and including) the third equality, we relied only on the i.i.d.\ nature of the noise. Moreover, the third equality provides a prescription on how to generalize: for a given choice of noise model, simply plug in the right cumulative function and probability distribution of that model. Thus, for a Gaussian noise for example, that is, $n_i\sim\mathcal{N}(\mu,\sigma^2)$, we have
\begin{equation}
p(z_k^i|z_{-k}) =\frac{1}{\sqrt{2\pi\sigma^2}}\int_{-\infty}^{\infty} d\epsilon \prod_{j\ne i}\Phi\left(\frac{\ell_i-\ell_j +\epsilon}{\sigma}\right)e^{-\frac{\epsilon^2}{2\sigma^2}} 
\end{equation}
where $\Phi$ is the normal cumulative distribution function. This is known as the independent probit model \citep{train2009discrete}. Of course, unlike the Gumbel case, in the Gaussian case this does not result in a closed form formula. This, however, does not prevent the GSP process from exploring the utility terrain of the model, given the functional similarity between Gumbel and normal distributions. 

If the noise cannot be assumed to be i.i.d., the third equality no longer holds. We see two main potential sources of i.i.d.\ violations in this paradigm:

\begin{enumerate}
    \item Intra-participant correlation (different participants have different utility functions);
    \item Neighboring-point correlation (neighboring points on the slider are likely to receive correlated noise).
\end{enumerate}

\begin{figure}[b]
  \centering
  \includegraphics[width=\linewidth]{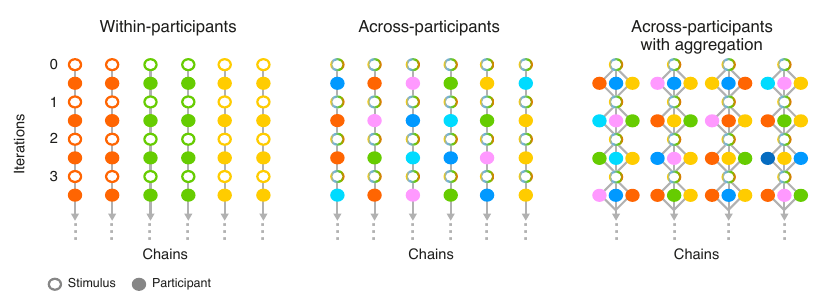}
  \caption{Illustration of different chain designs.}
  \label{fig:chain-types}
\end{figure}

Intra-participant correlation has different implications for across- and within-participant chains (Fig.\ \ref{fig:chain-types}). In across-participant chains, each participant only contributes one observation to the chain, so intra-participant correlation never manifests. In within-participant chains, all observations in a given chain come from the same participant, meaning that the i.i.d.\ assumption remains unviolated, and each chain ends up approximating the underlying utility function for the individual participant. The population-level utility function can then be approximated by aggregating over chains. 

Neighboring-point correlation could have a subtle effect on the derivations presented here. Future work could revise our model to include a correlation structure for these points, for example following the correlated multinomial probit model where noise values are taken from a joint Gaussian distribution with a specified correlation structure \citep{train2009discrete}.

Our derivation also does not take into account context effects, whereby the participant's previous trials influence their responses to the present trial. In particular, it is possible that the utility value for a given stimulus changes when the participant has already experienced that stimulus multiple times. This possibility is particularly high in within-participant chains, where the same participant experiences many stimuli from adjacent steps in the Gibbs sampler; in contrast, across-participant chains mostly avoid this effect by preventing the participant from experiencing multiple stimuli from the same chain (Fig.\ \ref{fig:chain-types}). We tested the strength of this effect in the emotional prosody experiment, conducting both a within-participants and an across-participants version of the same paradigm. We found that the results did not differ materially between the two, implying that memory effects were not a significant problem for this paradigm.


The above GSP derivation also assumes that the participant visits all slider positions. If this assumption is violated, the denominator in the GSP choice probability would cover only a subset of the locations, effectively reducing the granularity of the slider. We can try and minimize this effect experimentally, by forcing the participant to explore a certain amount of the stimulus space before proceeding to the next trial, but it is often impractical to enforce exhaustive exploration. However, we expect that the consequences of this assumption violation are not severe for two reasons: (a) participants tend to focus on the parts of the slider that contain most of the utility/probability mass, and (b) participants can extrapolate between slider locations to estimate the utility values of intermediate points. Nonetheless, we would like to explore this assumption more in future work.

\section{General methods}\label{appendix-general-methods}


\subsection{Implementation}

\begin{figure}
  \centering
  \includegraphics[width=\linewidth]{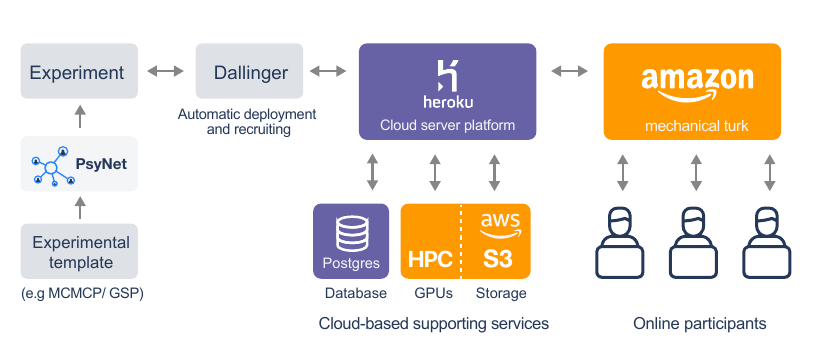}
  \caption{Computational infrastructure used for data collection.}
  \label{fig:architecture}
\end{figure}

We implemented all experiments in PsyNet, our under-development framework for implementing complex experiment paradigms such as GSP and MCMCP. This framework builds on the Dallinger platform for experiment hosting and deployment.\footnote{\url{https://github.com/Dallinger/Dallinger}} Participants interact with the experiment via their web browser, which communicates with a back-end Python server cluster responsible for organizing the timeline of the experiment (Fig.\ \ref{fig:architecture}). This cluster is mostly managed by Heroku,\footnote{\url{https://www.heroku.com/home}} and comprises a customizable collection of virtual instances that share the experiment management and stimulus generation workload, as well as an encrypted Postgres database instance for storing results. In some experiments we additionally used Amazon Web Services (AWS) S3 storage for hosting stimuli, and an AWS Elasic Compute Cloud (EC2) instance with an NVIDIA K80 GPU for deep neural network synthesis.\footnote{\url{https://aws.amazon.com/}}
Code for the implemented experiments can be found at \url{https://doi.org/10.17605/OSF.IO/RZK4S}.

\subsection{Participants}\label{appendix-participants}

All participants provided informed consent in accordance with the Max Planck Society Ethics Council approved protocol (application 2018-38). All participants were recruited from Amazon Mechanical Turk (AMT),\footnote{\url{https://www.mturk.com}} which is an online service for crowd-sourcing workers for online tasks. Here the only universal constraint we placed on recruitment was that participants must be at least 18 years of age, and have a 95\% or higher approval rate on previous tasks on AMT; this approval criterion is meant to help recruit reliable participants. In some experiments we also constrained the worker to be a US resident.

When designing the experiment, each component was given an estimate for the average time it should take to complete; participants were then paid at a US \$9/hour rate according to how much of the experiment they completed. Importantly, participants were still paid a proportional amount even if they left the experiment early on account of failing a pre-screening task.

A total of 5,178 participants took part in the 25 experiments reported in this paper, excluding those who failed pre-screening tests.
For the participants who reported demographic information, self-reported ages ranged from 18 to 89 (\emph{M} = 35.25, \emph{SD} = 10.37), and 35.74\% identified as female (63.9\% male and 0.36\% other).\footnote{Age and gender distributions were computed from all participants who passed the pre-screening tasks, excluding the validation experiments for emotional prosody, for which demographic information was not collected. Participant numbers only include participants who contributed at least one valid trial to the main experiment.}

These participants may be further differentiated into two groups: those who participated in the main experiments and those who participated in the validation experiments. These two groups had similar compositions, with the main participant group (Table S1) comprising 2,967 participants (35.95\% female, 63.64\% male, 0.41\% other; ages 18--89, \textit{M} = 35.25, \textit{SD} = 10.42), and the validation group (Table S2) comprising 2,211 participants (35.22\% female, 64.53\% male, 0.25\% other; ages 18--74, \emph{M} = 35.25, \emph{SD} = 10.24).

The musical chord study also collected additional information about musical expertise. Participants in the main chord experiment reported 0--25 (\emph{Med} = 2, \emph{M} = 4.26, \emph{SD} = 6.23) years of musical experience (i.e., playing an instrument or singing), whereas participants in the corresponding validation experiment reported 0--64 (\emph{Med} = 2, \emph{M} = 4.39, \emph{SD} = 7.57) years of musical experience. 

Participant recruitment was managed by PsyNet. For the across-participant chain experiments, we specified a desired number of chains and a desired length for these chains, and participants were then automatically recruited until the chains reached their desired lengths. For the within-participant chain experiments, we specified a desired number of completed participant sessions, and recruitment continued until this threshold was met. For the rating experiments, we chose a desired number of ratings per experimental condition\footnote{An experimental condition typically corresponded to one point on a figure, for example the third iteration for the second `lavender' GSP chain.} such that we expected that any variation in the resulting condition means should primarily reflect the stochasticity of the original sampler rather the stochasticity of participant raters. As a rule of thumb, we aimed for approximately 150 participants per validation study, scaling this number accordingly when the validation study compared multiple methods. Participants were then automatically recruited until the minimum number of ratings per experimental condition was reached.


\begin{landscape}
\begin{table}
  \caption{Main experiments}
  \label{experiments-table}
  \centering
  \begin{tabular}{llcccclclcl}
    \toprule
    Experiment & Method & Rep. & Dim. & Iter. & Agg. & Chain type
    & $N$ & Pre-screening & US-only & Validated in \\
    \cmidrule(r){1-11}
    1a Color (MCMCP) & MCMCP & 8  & 3 & 30 & 1 & Across & 57 & CB, CV & No & Exp.\ 1d, 1h \\
    1b Color (GSP) & GSP & 8  & 3 & 30 & 1 & Across & 53 & CB, CV & No & Exp.\ 1d, 1h     \\
    1c Color (agg.\ GSP) & GSP & 8  & 3 & 30 &  10 & Across & 312 & CB, CV & No & Exp.\ 1d, 1h  \\
    1e Color (MCMCP proposal) & MCMCP  & 8  & 3 & 30 & 1 & Across & 153 & CB, CV & No & - \\
    1f Color (questions) & GSP/MCMCP  & 8  & 3 & 30 & 1 & Across & 190 & CB, CV & No & - \\
    1g Color (agg.\ MCMCP) & MCMCP & 8  & 3 & 30 & 10 & Across & 302 & CB, CV & No & Exp.\ 1h \\
    2a Prosody (within) & GSP & 3  & 7 & 21 & 1 & Within & 110 & Audio & Yes & Exp.\ 2c, 2d  \\
    2b Prosody (across) & GSP & 3  & 7 & 20 & 1 & Across & 57  & Audio & Yes & Exp.\ 2d \\
    3a Musical chords & GSP  & 1  & 2 & 40 & 1 & Across & 134  & Audio & No & Exp.\ 3b \\
    4a Faces & GSP & 6  & 10 & 50 & 5 & Across & 293 & CV & Yes & Exp.\ 4b, 4d, 4g \\
    4c Faces (KDE) & GSP & 6  & 10 & 50 & 5 & Across & 278 & CV & Yes & Exp.\ 4d \\
    4e Faces (basis) & GSP & 1  & 10 & 30 & 5 & Across & 167 & CV & Yes & Exp.\ 4f \\
    4h Faces (art) & GSP & 6  & 10 & 50 & 5 & Across & 260 & CV & Yes & - \\
    4i Faces (global, KDE) & GSP & 6  & 10 & 50 & 5 & Across & 269 & CV & No & Exp.\ 4d \\
    4j Faces (dating) & GSP & 1  & 10 & 30 & 5 & Across & 332 & CV & Yes & - \\
    \bottomrule
  \end{tabular}
  \begin{tablenotes}
    \item $Note.$ `Rep.' indicates the number of semantic representations that were tested; `Dim.' indicates the dimensionality of the stimulus space; `Iter.' indicates the number of iterations in each chain; `Agg.' indicates how many participants contributed to each iteration of the GSP chain; `$N$' denotes the number of participants included in the final analysis; `CB' denotes the color blindness pre-screening task; `CV' denotes the color vocabulary pre-screening task; `US-only' indicates whether the participant group was restricted to US residents; `Exp.' denotes `Experiment'.
\end{tablenotes}
\end{table}

\begin{table}
  \caption{Validation experiments}
  \label{validation-experiments}
  \centering
  \begin{tabular}{lcccclcl}
    \toprule
    Experiment & \vtop{\hbox{\strut Ratings per}\hbox{\strut participant}} & \vtop{\hbox{\strut Ratings per}\hbox{\strut stimulus}} & Total stimuli
    & $N$ & Pre-screening & US-only & Validating \\
    \cmidrule(r){1-8}
    1d Color (original) & 60 & 5.2 & 3,720 & 322 & CB, CV & No & Exp.\ 1a, 1b, 1c \\
    1h Color (inc.\ agg.\ MCMCP) & 60 & 3.3 & 4,960 & 270 & CB, CV & No & Exp.\ 1a, 1b, 1c, 1g \\
    1i Color (uniform sample) & 60 & 4.2 & 4,000 & 280 & CB, CV & No & Exp.\ 1a, 1b, 1c, 1g \\
    2c Prosody  & 147 & 5.4 & 4,383 & 161 & Audio & Yes & Exp.\ 2a  \\
    2d Prosody & 132 & 4.1 & 4,874 & 153 & Audio & Yes & Exp.\ 2a, 2b \\
    3b Musical chords & 80 & 16.4  & 820 & 168 & Audio & Yes & Exp.\ 3a \\
    4b Faces (original) & 80 & 52.1  & 275 & 179 & CV & Yes & Exp.\ 4a  \\
    4d Faces (aggregation, location) & 80 & 25.6  & 815 & 261 & CV & No & Exp.\ 4a, 4c, 4i  \\
    4f Faces (basis) & 59.9 & 4.3 & 260 & 131 & CV & No & Exp.\ 4e  \\
    4g Faces (bias) & 78.9 & 3.2  & 7,056 & 286 & CV & Yes & Exp.\ 4a  \\
    \bottomrule
  \end{tabular}
  \begin{tablenotes}
    \item $Note.$ `$N$' denotes the number of participants included in the analysis; `CB' denotes the color blindness pre-screening task; `CV' denotes the color vocabulary pre-screening task; `US-only' indicates whether the participant group was restricted to US residents; `Exp.' denotes `Experiment'. In the row corresponding to Exp.\ 4g, the number of stimuli (7,056) corresponds to 7 (the number of questions) multiplied by 1,008 (the number of images). 
\end{tablenotes}
\end{table}
\end{landscape}

\subsection{Pre-screening tests}\label{appendix-prescreening}

A useful technique for improving the quality of data from online participants is to implement pre-screening tests designed to screen out participants likely to deliver low-quality data \citep{woods2017headphone}. 
Here we used three pre-screening tests in various combinations: 
a color blindness test, a color vocabulary test, and an audio test.
These tests are primarily intended to screen out participants who do not meet certain explicit criteria such as wearing headphones, 
but they also help to screen out participants with a minimal degree of English comprehension, 
or automated scripts (`bots') masquerading as participants \citep{chmielewski2020mturk}.

\begin{figure}
  \centering
  \frame{\includegraphics[width=0.5\linewidth]{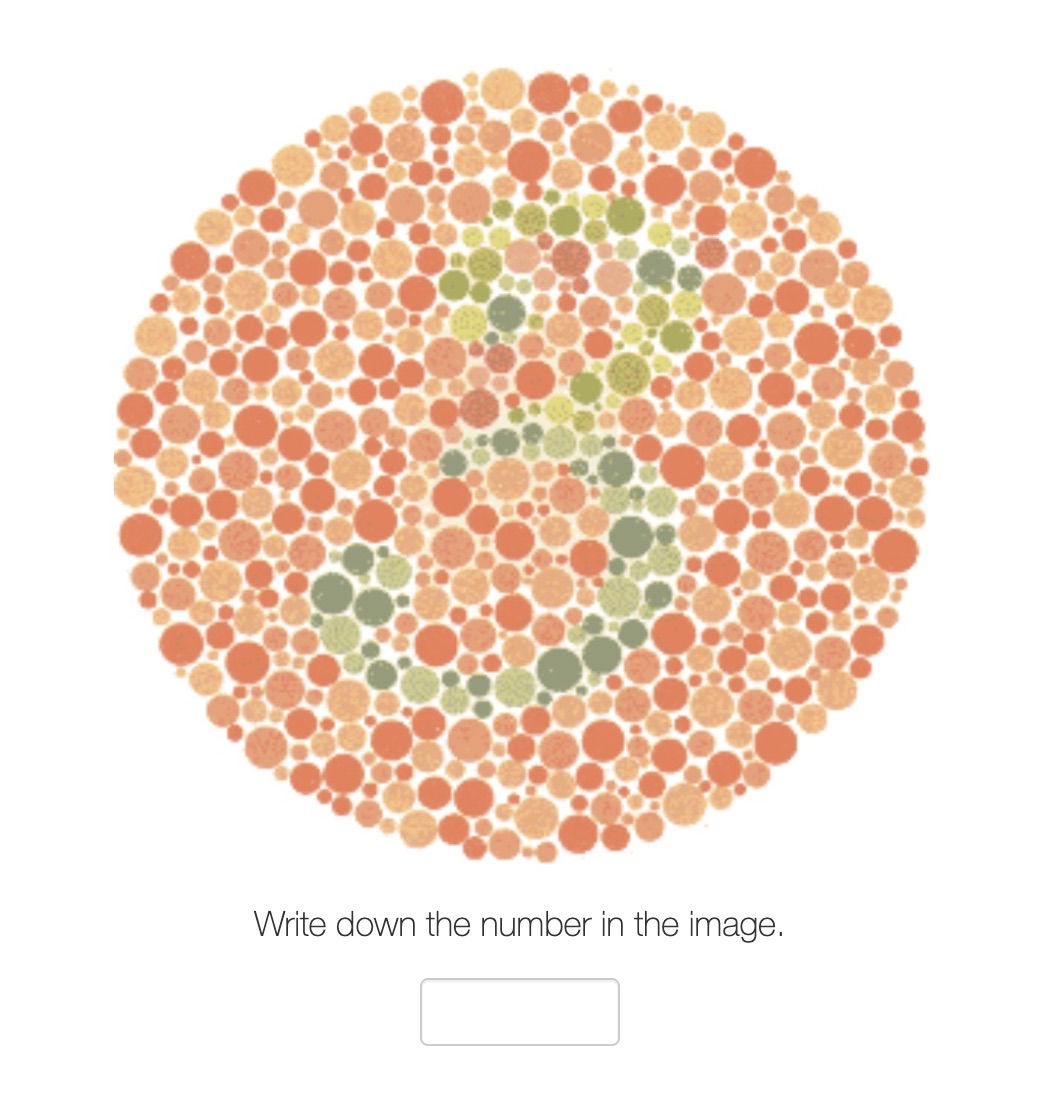}}
  \caption{Example trial from the color blindness pre-screening task.}
  \label{fig:color-blindness}
\end{figure}

The color blindness test was derived from the well-known Ishihara color blindness test \citep{colorblind}. Here participants had to respond to six trials where the task was to transcribe a number from an image, with the contrast of the image being designed such that it is difficult to perform if the participant suffers from color perception deficiencies (see Fig.\ \ref{fig:color-blindness} for an example trial). The image was set to disappear after three seconds to encourage quick responses. Each participant had to take six such trials; to pass, they had to answer at least four of these six trials correctly.

\begin{figure}
  \centering
  \frame{\includegraphics[width=0.75\linewidth]{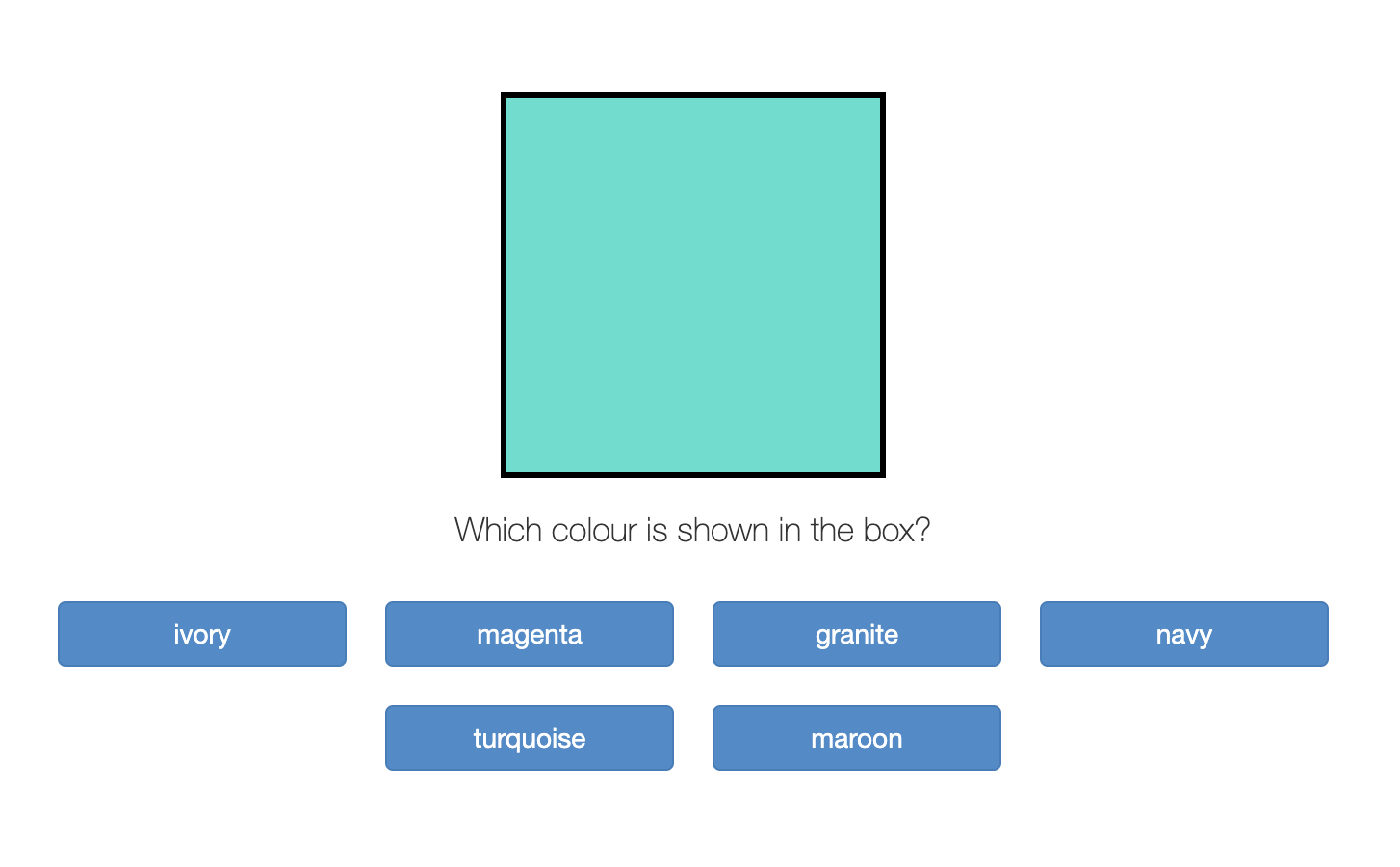}}
  \caption{Example trial from the color vocabulary pre-screening task.}
  \label{fig:color-vocab}
\end{figure}

The color vocabulary test was constructed by taking six English color words that require a relatively good vocabulary knowledge to understand: `turquoise', `magenta', `granite', `ivory', `maroon', and `navy'. None of these words were used in the other experiments. We associated each word with an RGB definition sourced from Wikipedia, and presented the participant with six trials where they were presented with a color and had to choose which of the six words corresponded to that color (see Fig.\ \ref{fig:color-vocab}). The pass threshold was a score of four out of six.

The audio pre-screening task, originally developed in \cite{woods2017headphone}, was intended to ensure that participants were wearing headphones and could hear perceive subtle sound differences. The task has participants perform a three-alternative forced-choice task to identify the quietest of three tones. These tones are constructed to elicit a phase cancellation effect, such that when played on loudspeakers the order of quietness changes, causing the participant to respond incorrectly. Each participant had to take six such trials; to pass, they had to answer at least four of these six trials correctly.

\begin{figure}
  \centering
  \frame{\includegraphics[width=0.75\linewidth]{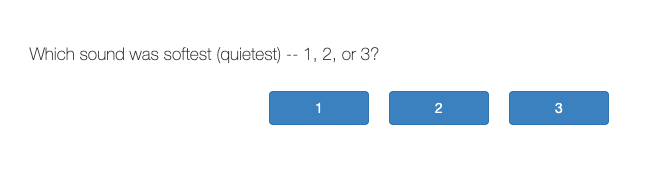}}
  \caption{Example trial from the audio pre-screening task.}
  \label{fig:sound-check}
\end{figure}

\subsection{Performance incentives}

In order to further improve data quality, some of our experiments (specifically, all but the emotional prosody experiments) additionally included a financial incentive for participants to provide high-quality data.
Prior to the main part of the experiment, we informed all participants of this incentive, using the following text:

\begin{quote}
    The quality of your responses will be automatically monitored,
    and you will receive a bonus at the end of the experiment
    in proportion to your quality score. The best way to achieve
    a high score is to concentrate and give each trial your best attempt.    
\end{quote}

We purposefully left the definition of `quality' vague, so as to avoid encouraging participants to `game' a particular aspect of response quality. Of course, our tasks were subjective, and so there was no meaningful way to define a high-quality answer \textit{a priori}. Instead, our approach was to use consistency as a proxy for quality; the rationale is that a participant who takes the task seriously and carefully is likely to deliver consistent responses when administered the same trial multiple times, in contrast to a participant who does not pay attention to the task and simply answers randomly. 

We estimated consistency as follows. Once a participant finished all of their `main' experiment trials, they then received a small number 
(4--8, depending on the experiment)
of trials that repeated randomly selected trials from the earlier part of the experiment. The data from these trials contributed solely to consistency estimation, not to chain construction. In GSP trials and four-point rating trials, consistency was quantified by taking the Spearman correlation between the two sets of answers; for MCMCP trials, consistency was quantified by taking the percentage agreement between the two sets of answers. Participants were then given a small monetary bonus in proportion to the resulting consistency score, ranging from zero dollars for chance performance up to one dollar for perfectly consistent performance.


\subsection{Chain construction}\label{appendix-chains}

In all experiments except the emotional prosody experiment, we randomized the starting locations of each chain by randomly sampling from a uniform distribution over the range of permissible feature values. In the case of emotional prosody, we found this randomization problematic because it often led to unrealistic parts of the stimulus space. In this case we therefore initialized each chain at a `null' state corresponding to the unaltered reference sentence.

In a given trial of a GSP experiment, the participant's slider manipulated exactly one dimension of the stimulus. To counteract any potential biases towards left or right slider directions, we randomized the effective direction of the slider on each trial, such that approximately half of the time the right of the slider corresponded to positive feature values, and the other half of the time it corresponded to negative feature values.

Our experiments implemented both within-participant and across-participant chains
(Fig.\ \ref{fig:chain-types}).
In {within-participant} chains, the entire chain is completed by just one participant, and the resulting samples reflect the semantic representations of that single participant. In {across-participant} chains, each iteration comes from a different participant, and the samples then reflect shared semantic representations across participants.

Across-participant chains are more complex to implement because of the interaction between multiple participants. Each time a participant is ready to take a new trial, it is necessary to scan the different chains in the experiment and identify one that satisfies the following conditions: 

\begin{enumerate}
    \item The chain is not full (i.e., it has not reached its specified quota of iterations);
    \item The participant has not already participated in that chain;
    \item No other participants have been assigned to that particular iteration of the chain.
\end{enumerate}

The last point -- ensuring that multiple participants are not assigned to the same iteration of a chain -- is important for the efficiency of data collection, but it can cause problems when a participant claims a particular iteration of the chain and then drops out of the experiment, potentially blocking any future additions to that chain. We therefore implemented a time-out parameter for this experiment, set to 60 seconds, after which the participant's pending trial was invalidated and the chain was unblocked.

In within-participant chains we are free to discard all of a participant's data when they drop out of an experiment partway through. This is not practical in across-participant chains, however, where many subsequent participants might have built on the data previously contributed by this participant. In the latter case, we therefore retain the participant's contributions even when they drop out of the experiment.

\section{Color}\label{appendix-color}


\subsection{Supplementary methods}\label{appendix-color-methods}

We chose eight words designed to be moderately but not overly familiar to English speakers that we anticipated to evoke strong color associations. These words were `chocolate', `cloud', `eggshell', `grass', `lavender', `lemon', `strawberry', and `sunset'. 
We then explored the perceptual spaces associated with these eight words using GSP and MCMCP.

We implemented GSP and MCMCP using the Hue, Saturation, Lightness (HSL) color space. We chose this color space over the Red, Green, Blue (RGB) color space because it is generally considered to better reflect how humans perceive color relationships. In this space, each color is encoded as three integers: hue, saturation, and lightness, taking values in [0, 360), [0, 100], and [0, 100] respectively. 

MCMCP relies on the specification of a proposal function. In our main experiment, we used a Gaussian
distribution with a standard deviation of 30, chosen on the basis of internal piloting, and rounding samples to the nearest integer values.
In MCMCP the proposal distribution should be symmetric, which can be problematic to satisfy when the sampler reaches the boundaries of the sample space. We addressed this problem by computing the proposal distribution modulo the scale range, such that moving past the top of the scale means returning to the bottom of the scale. This works particularly well for hue, which is already defined as a circular space, with both 0 and 360 corresponding to the color red. It works less well for saturation and lightness, because these are linear scales; however, as we chose target colors occupying central regions of these two scales, we expected that these boundary effects should not materially influence MCMCP's performance.

\begin{figure}
  \centering
  \frame{\includegraphics[width=0.65\linewidth]{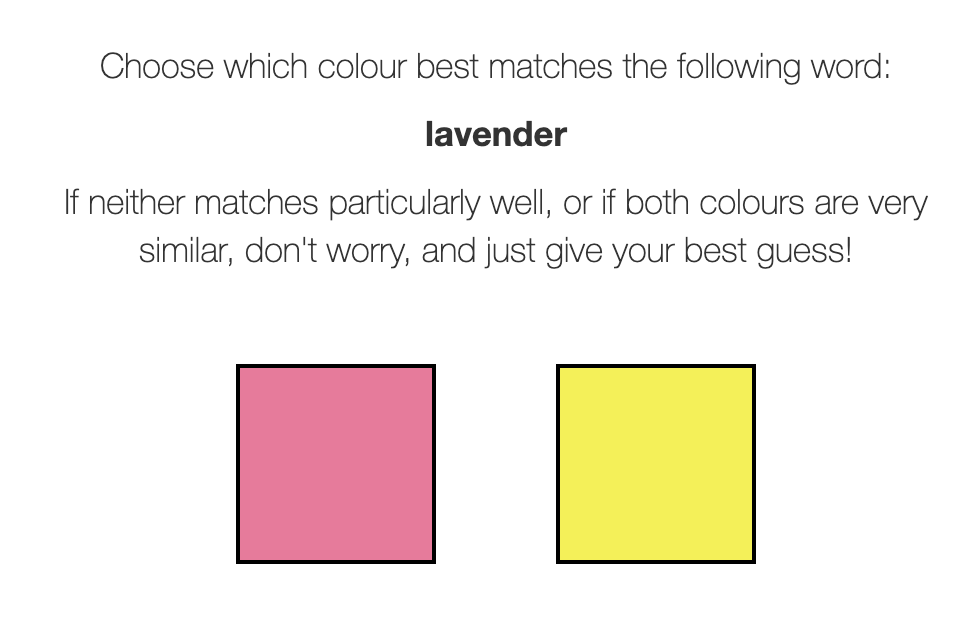}}
  \caption{Screenshot from the color MCMCP implementation.}
  \label{fig:color-mcmcp-trial}
\end{figure}

\begin{figure}
  \centering
  \frame{\includegraphics[width=0.65\linewidth]{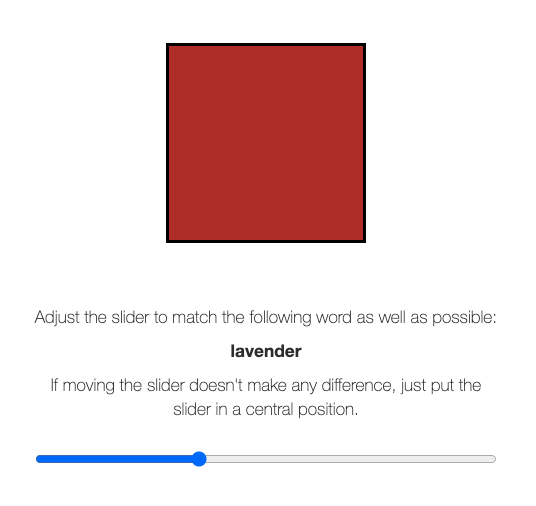}}
  \caption{Screenshot from the color GSP implementation.}
  \label{fig:color-gsp-trial}
\end{figure}

We implemented simple web interfaces for the MCMCP and GSP tasks. In the MCMCP task, participants were presented with pairs of colors, and had to choose which color best represented a target word (Fig.\ \ref{fig:color-mcmcp-trial}). In the GSP task, participants were presented with a single color that constantly updated to reflect the current position of a slider; participants were then instructed to move the slider to make the color represent a target word as well as possible (Fig.\ \ref{fig:color-gsp-trial}).

For each given sampling method we constructed five across-participants chains per adjective, yielding 40 chains in total. Each chain was filled to a length of 30 states, not including the initial random state. Each participant contributed a maximum of 40 trials to the chains for a given sampling method
(Exp.\ 1a--c; see Table \ref{experiments-table} for participant numbers). 

The aggregated GSP experiment combined 10 trials for each iteration of the Gibbs sampler (Exp.\ 1c). These 10 trials were combined using the arithmetic mean in the case of saturation and lightness, and the circular mean in the case of hue. These means were then propagated to the next iteration of the Gibbs sampler. 

We also ran five follow-up experiments to better understand the relative performance of GSP and MCMCP
(see also Tables \ref{experiments-table} and \ref{validation-experiments}):

\begin{itemize}
    \item We tested MCMCP with five different proposal function standard deviations: 10, 20, 30, 40, and 50, all expressed on the integer color scale (Exp.\ 1e).
    \item We reran the MCMCP and GSP experiments with three different kinds of questions designed to probe different notions of utility and category membership (Table \ref{tab:questions}, Exp.\ 1f).
    \item We reran the MCMCP experiment using 10-fold aggregation (Exp.\ 1g), and validated it alongside the other methods (Exp.\ 1h).
    \item We tested 4,000 colors randomly sampled from a uniform distribution over the HSL space using the same rating procedure as Exp.\ 1d (Exp.\ 1i).
\end{itemize}

\begin{table}
\caption{Questions used in Exp.\ 1f.}
\label{tab:questions}
\begin{tabular}{lp{5.5cm}p{5.5cm}}
 \toprule
 Label & MCMCP & GSP \\
 \cmidrule(r){1-3}
Probability &
  Choose which colour is most likely to come from the following category. &
  Adjust the slider to make the color as likely as possible to come from the following category. \\
Goodness &
  Choose which colour best matches the following word. &
  Adjust the slider to match the following word as well as possible. \\
Typicality &
  Choose which colour is most typical of the following category. &
  Adjust the slider to make the color as typical as possible for the following category. \\
  \bottomrule
\end{tabular}
\begin{tablenotes}
    \item $Note.$ All color experiments except for Exp.\ 1f used only the `Goodness' question; Exp.\ 1f tested all three questions.
\end{tablenotes}
\end{table}

The validation experiment (Exp.\ 1d) used the same pre-screening procedure as the chain-construction experiments. A minimum of five ratings were collected for each sample generated in the former experiments, with the constraint that participants could not rate the same stimulus more than once. Participants were assigned pseudo-randomly to stimuli such that the number of ratings accumulated evenly for each stimulus. In each trial, participants were presented with the target word from the original chain, and asked to judge how well the colour matched this word on a scale from 1 (not at all) to 4 (very much). A given participant's ratings were only included in the final tallies if they completed the entire validation experiment.
See Table \ref{validation-experiments} for participant numbers.

\subsection{Supplementary results} \label{appendix-color-results}

In the main paper we identified a clear advantage for GSP over MCMCP, given chains of the same length and the same amount of aggregation. However, we were concerned about several possible confounds, which we will now discuss alongside corresponding analyses.

\begin{figure}
  \centering
  \includegraphics[width=0.6\linewidth]{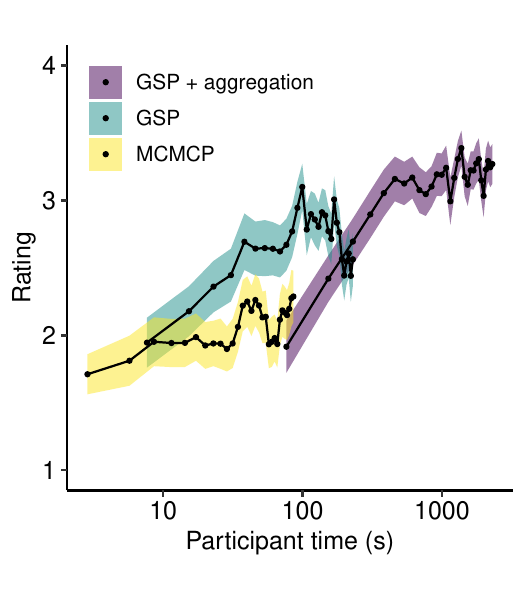}
  \caption{Mean sample ratings as a function of the participant time invested in chain construction (Exp.\ 1a--c, 1d), with time plotted on a log scale (95\% confidence intervals over participants).}
  \label{fig:color-validation-profile-normed}
\end{figure}

\textbf{\textit{Claim:} GSP trials are more time-consuming than MCMCP trials. Even if GSP requires fewer trials to achieve good sample quality, if these trials take much longer, then GSP will end up being practically slower than MCMCP.}
Fig.\ \ref{fig:color-validation-profile-normed} plots validation ratings for the different sampling methods, with the horizontal axis now corresponding to the total participant time invested in the respective chains (Exp.\ 1a--c). 
We estimated total participant time by taking iteration number and multiplying it by the median participant time spent on the two different trial types. 
The results indicate that non-aggregated GSP still clearly outperformed MCMCP despite the longer duration of its individual trials. 
It is difficult to make a clear statement about the relative performance of aggregated GSP because its profile overlaps minimally with the other two methods; however, the figure implies that non-aggregated GSP outperforms aggregated GSP for the first few iterations, with aggregated GSP then overtaking at a later point. 
This is consistent with our expectations: the fast-but-noisy non-aggregated GSP can quickly escape its low-probability starting states, but the same noise prevents it from converging as precisely as aggregated GSP in later iterations.

\begin{figure}
  \centering
  \includegraphics[width=1.0\linewidth]{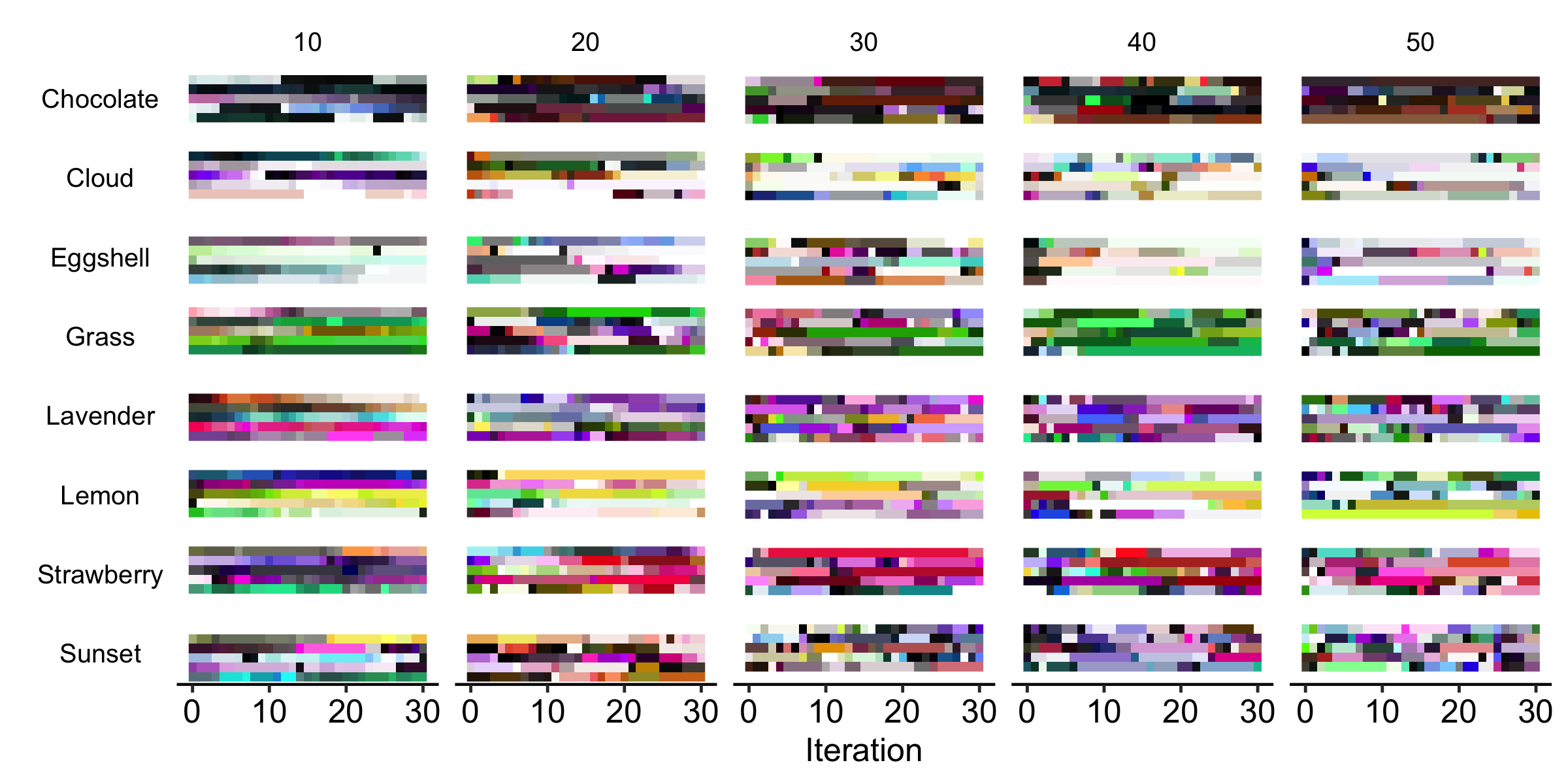}
  \caption{Raw color samples for MCMCP with five different standard deviations for the Gaussian proposal function: 10, 20, 30, 40, and 50 (Exp.\ 1e).}
  \label{fig:mcmcp-explore}
\end{figure}

\textbf{\textit{Claim:} MCMCP has a tuning parameter corresponding to the width of the proposal function. Perhaps the relatively poor performance of MCMCP was simply due to the wrong choice of proposal width.} We tabulated samples from the previously described control experiment with different MCMCP proposal widths (Fig.\ \ref{fig:mcmcp-explore}, Exp.\ 1d). There appears to be little difference in sample quality for the different proposal widths.
As expected, we see that the MCMCP chains with the smallest proposal width (10) only make local adjustments to the color, meaning that once the chain gets close to an appropriate color category, it can be carefully tweaked to resemble this category as well as possible. However, these narrow-proposal chains often fail to approach the appropriate color category in the first place, even after 30 iterations. In contrast, the wide-proposal chains explore the color space quickly, but are unable to make subtle adjustments to match specific categories. The moderate proposal width of 30 provides some compromise between these two behaviors, and seems to be a sensible choice for the MCMCP-GSP comparison.

\begin{figure}
  \centering
  \includegraphics[width=1.0\linewidth]{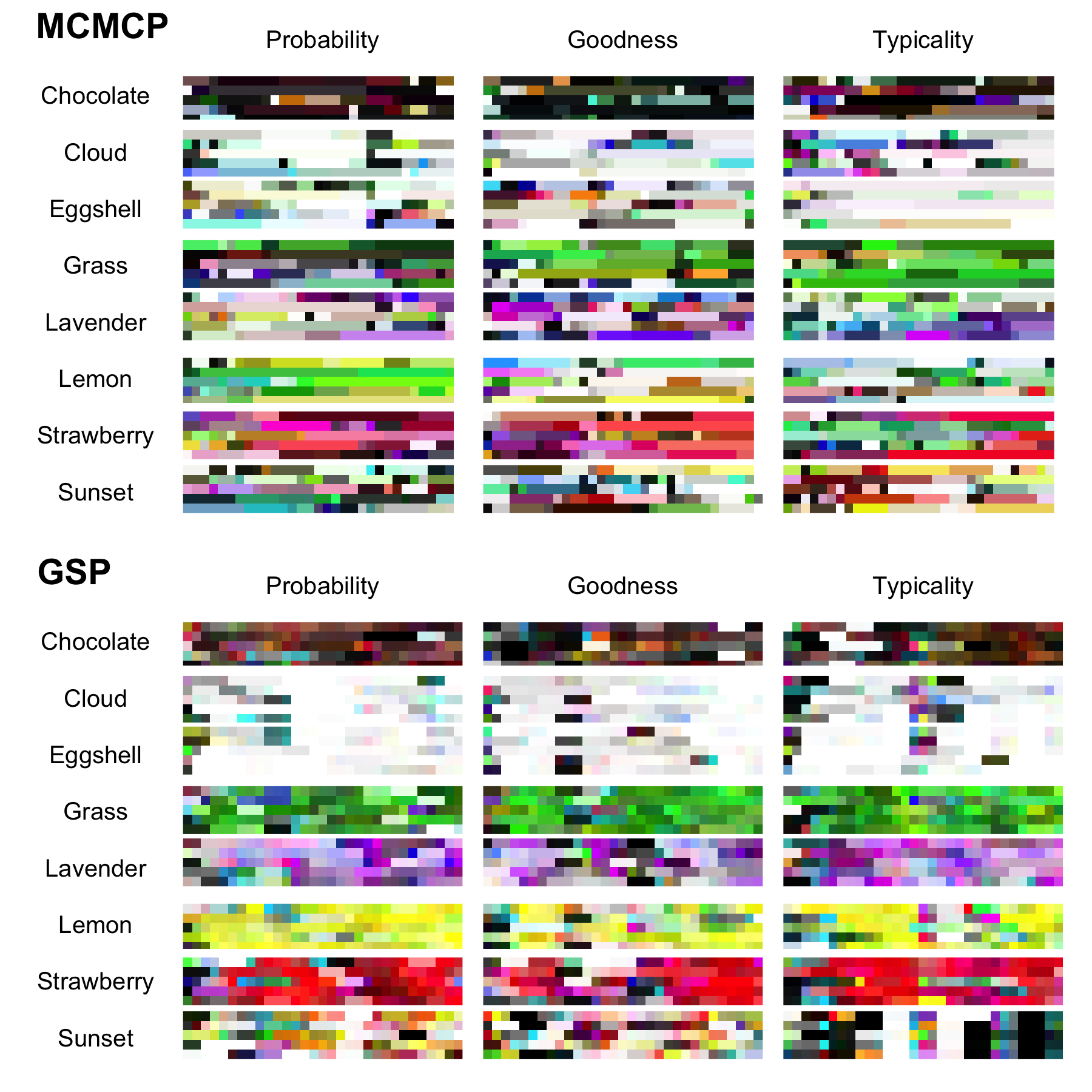}
  \caption{Raw color samples for MCMCP and GSP with three different kinds of questions as described in Table \ref{tab:questions} (Exp.\ 1f).}
  \label{fig:color-questions}
\end{figure}

\textbf{\textit{Claim:} We altered the MCMCP question somewhat to better represent the notion of continuous utility as opposed to category membership. Perhaps this alteration diminished the efficacy of MCMCP in practice.}
We tabulated samples from Exp.\ 1f which trialled different types of questions for the MCMCP and GSP tasks (Table \ref{tab:questions}, Fig.\ \ref{fig:color-questions}). Visually inspecting these plots, we struggled to discern any systematic effect of question type on the sample distributions. We do not doubt that subtle differences could be distinguished with the right kind of experiment, but it seems that in practice any such effects are small.

\begin{figure}
  \centering
  \includegraphics[width=0.7\linewidth]{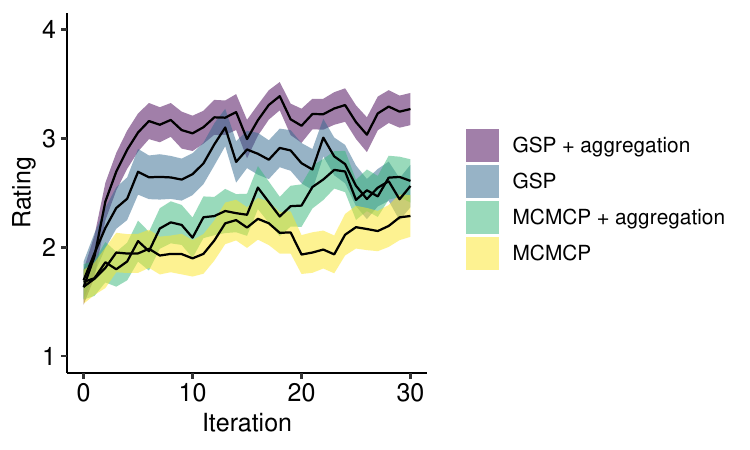}
  \caption{Validation results for non-aggregated and aggregated GSP and MCMCP (Exp.\ 1a, 1b, 1c, 1g, and 1h). The shaded regions indicate 95\% confidence intervals over participants.}
  \label{fig:color-supplementary-validation}
\end{figure}

\textbf{\textit{Claim:} We only evaluated MCMCP without aggregation; perhaps MCMCP with aggregation would perform as well as GSP.}
We compared validation ratings for aggregated MCMCP against ratings for aggregated GSP, non-aggregated GSP, and non-aggregated MCMCP (Fig.\ \ref{fig:color-supplementary-validation}). Aggregated MCMCP does outperform non-aggregated MCMCP, but the difference is small compared to the difference between GSP and aggregated GSP. This makes intuitive sense: while aggregated GSP can produce very precise updates at each iteration, aggregated MCMCP can only provide one bit of information at each iteration, placing a fundamental limit on its convergence rate.

\begin{figure}
  \centering
  \includegraphics[width=0.8\linewidth]{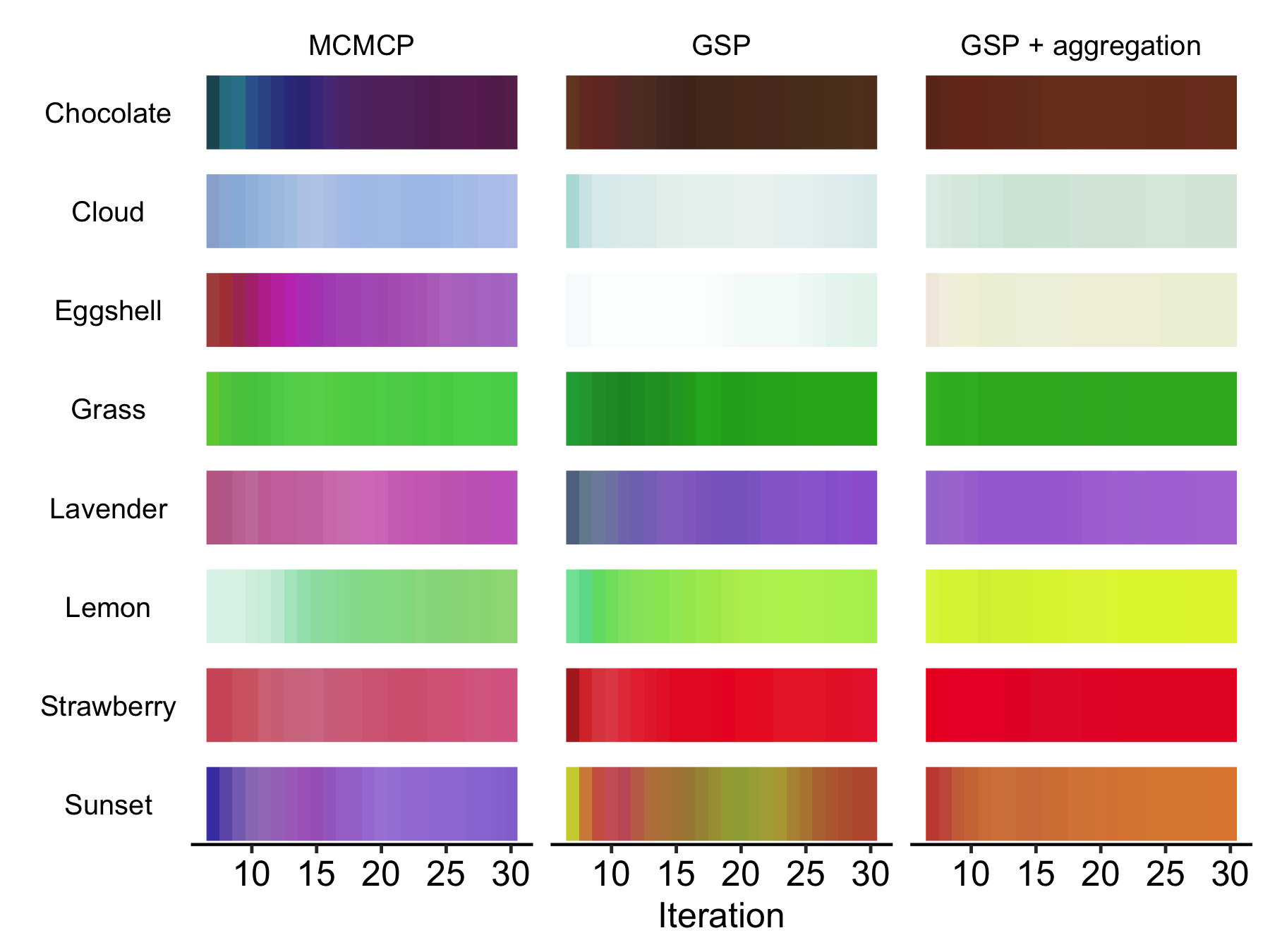}
  \caption{Colors derived by averaging raw samples from iteration 7 onwards for the different sampling methods (Exp.\ 1a--c).}
  \label{fig:cumulative-color}
\end{figure}

\textbf{\textit{Claim:} A common analysis approach with MCMCP is to generate category prototypes by averaging over many samples. Perhaps MCMCP performs better when using this analysis method.} We recomputed the samples generated by the three methods using instead an incremental aggregation process, generating a summary sample for each iteration and target word by averaging all previous samples from all chains for that word, with iterations 1--6 treated as burn-in samples and hence discarded. The resulting samples are displayed in Fig.\ \ref{fig:cumulative-color}. The aggregation process clearly improves sample quality for MCMCP and GSP (non-aggregated), but it does not fully solve MCMCP's problem with poor sample quality. Though we do not have participant rating data for these aggregated samples, it is clear that MCMCP failed to converge on appropriate colors for chocolate, eggshell, and lavender. One might further criticize the lavender samples for being too red, the strawberry and sunset samples for not being red enough, and the lemon samples for being not yellow enough. It seems apparent that aggregating over trials does not necessarily resolve the performance issues of MCMCP in this paradigm.

\begin{figure}
  \centering
  \includegraphics[width=1.0\linewidth]{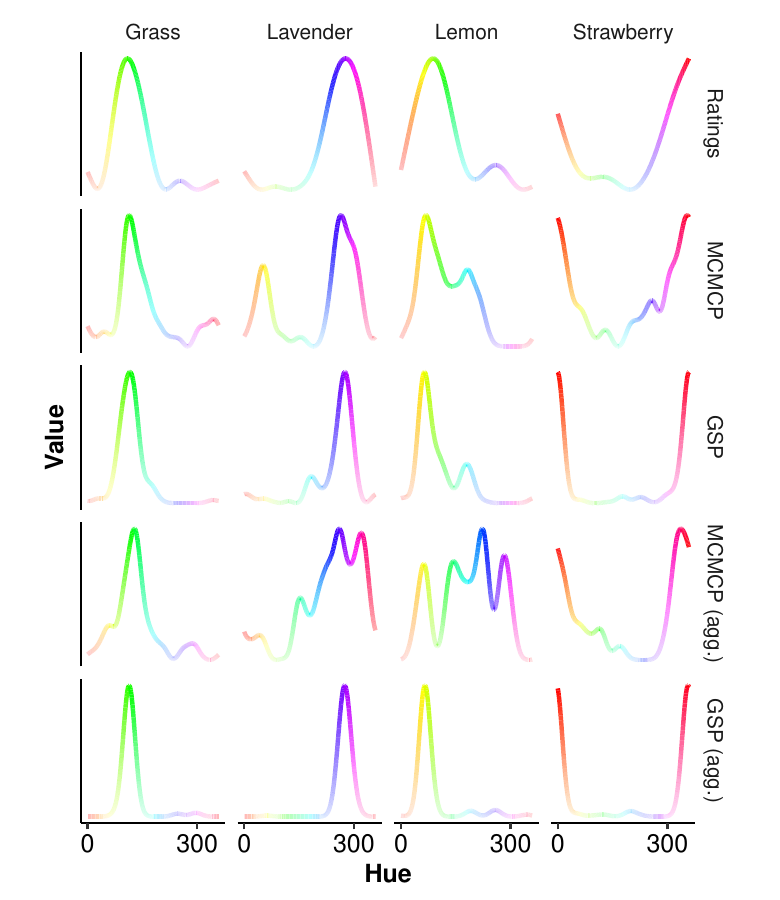}
  \caption{Utility distributions as estimated by rating, MCMCP, and GSP experiments, treating iterations 1--6 as burn-in samples (Exp.\ 1a, 1b, 1c, 1g, 1i).}
  \label{fig:color-1d-kernels}
\end{figure}

\textbf{\textit{Claim:} Our original evaluation rewards methods that produce highly prototypical category exemplars; using our utility function metaphor, one might say that the evaluation rewards mode-seeking behavior. 
However, there is a trade-off between mode seeking and exploration; perhaps GSP is better at mode seeking, but MCMCP is better at exploration.}
We estimated a benchmark utility distribution over the stimulus space for each target word using a large-scale rating experiment (Exp.\ 1i), and then compared the results to the utility distributions estimated by MCMCP and GSP. 
To provide a visual intuition for the differences between techniques, Fig.\ \ref{fig:color-1d-kernels} plots marginal distributions for hue as estimated by the rating, MCMCP, and GSP experiments, using a generalized additive model for the ratings and a kernel density estimator (KDE) for the MCMCP and GSP distributions, and again treating iterations 1--6 as burn-in samples. 
Only `grass', `lavender', `lemon', and `strawberry' are plotted here, because these are the four words with the most interpretable marginals for hue (the remaining adjectives have many very dark or very light samples, in which case differences in hue become imperceptible). 
From comparing GSP and MCMCP to the ratings, it is apparent that the poor performance of MCMCP is not simply due to having broader peaks, but rather comes from mislocated secondary peaks, for example red for `grass', orange for `lavender', blue for `lemon', and so on. 
Visually inspecting the raw samples in Fig.\ \ref{fig:color}B supports this impression: many of the MCMCP samples seem to be unrelated to the target category. 
Incidentally, the figure also helps for visualizing the effect of aggregation; we see how aggregation sharpens the GSP peaks to clear unimodal distributions, but fails to provide much improvement for MCMCP.
Future work should investigate these differences more systematically using quantitative assessments of multidimensional distribution similarity, and unpacking the potentially non-trivial relationship between sample ratings and utility values.

To summarize, it seems that none of these six considerations impact substantially on the main conclusion that GSP outperforms MCMCP for this color estimation task. Nonetheless, each of these issues could certainly be explored in more detail in future work; each perceptual domain is different, and in some cases MCMCP may become the preferred tool as a result.


\section{Emotional prosody}\label{appendix-emotion}


\subsection{Stimuli}\label{appendix-emotion-stimuli}

The stimuli were created on the basis of three sentences from the Harvard sentences \cite{ieee} recorded by a female speaker \cite{demonte2019}. These sentences are phonologically balanced and semantically neutral. The stimulus space was then defined through seven continuous acoustic manipulations performed to these sentences. The manipulations were performed using the software Praat \cite{praat} and the Python package Parselmouth \cite{parselmouth}.  Pitch (F0 contour) was extracted using a pitch floor of 100 Hz and ceiling of 500 Hz (default window size) using the command \texttt{To Pitch} in Parselmouth. Before proceeding we confirmed that all contours were free of any octave jumps. From the \texttt{Sound} and the \texttt{Pitch} object, we created a \texttt{Manipulation} object using the command \texttt{To Manipulation}. From the \texttt{Pitch} object we extracted the glottal pulses using \texttt{To PointProcess}. The manipulations were then performed in the following order:

\begin{enumerate}
    \item \textit{Pitch level}, shifting the pitch contour by a value in the range [$-$37, 37] Hz.
    
    \item \textit{Pitch range}, scaling the original pitch range (expressed in Hz) by a value in the range [20, 180]\%, using the middle of the original pitch range as the center of the scaling operation.
    
    \item \textit{Pitch slope}, altering the original sentence's pitch slope by a value in the range [$-$37, 37] Hz. In our case, the reference sentences always began with a falling slope, and our manipulation was never severe enough to change them to a rising slope. Instead, a positive value of our pitch slope feature indicates a flattened contour, and a negative value indicates a steeply falling contour.
    
    
    We manipulated pitch slope in the following way. We extracted the time of the first ($t_0$) and last ($t_1$) pitch values (ignoring unvoiced segments), and then edited the pitch contour by adding the following linear function $f(t)$ to each pitch:
    
    \begin{equation}
        f(t) = x*\frac{t-t_0}{t_1-t_0}
    \end{equation}
    
    where $t$ denotes the time of the point being edited and $x$ denotes the feature value, ranging between $-$37 Hz and 37 Hz.
    
    
    We achieved this by creating an empty \texttt{PitchTier} object and populating it with the new contour using the command \texttt{Add point}. Finally we replace the old \texttt{PitchTier} in the \texttt{Manipulation} object with the new one using \texttt{Replace pitch tier}.
    
    \item \textit{F0 perturbation} is commonly measured as local frequency variation in the F0 contour (jitter), and corresponds approximately to the perceptual impression of hoarseness \cite{Titze1987}. We modified F0 perturbation by converting the \texttt{PointProcess} object (representing the glottal pulses) to a Praat \texttt{Matrix} object (representing the time points of the pulses) using \texttt{To Matrix}. We changed the position of the pulses by applying the Praat formula \texttt{self + randomGauss(0, r)} where \texttt{r} was a number between 0 and 0.0001 determining the strength of the perturbation. The \texttt{Matrix} was converted back to a \texttt{PointProcess} with \texttt{To PointProcess}, and the glottal pulses in the \texttt{Manipulation} replaced using \texttt{Replace pulses}. This follows the algorithm proposed in \cite{vocaltoolkit}.
    
    \item \textit{Duration}, allowed to change linearly from 80\% to 120\% from the original duration. To manipulate the duration we created an empty \texttt{DurationTier} object using the command \texttt{Create DurationTier}. At time 0 we placed a point with the duration value using the command \texttt{Add point 0 scalar}. We then ran \texttt{Replace duration tier} to apply the changes. Note that changing the duration did not affect the overall pitch.
    
    \item \textit{Intensity variation}, corresponding to a periodic amplitude modulation of the signal. 
    This manipulation was characterized by two parameters which constituted two independent dimensions of the stimulus space: 
    \textit{amplitude modulation frequency} (ranging from 0--5 Hz) 
    and \textit{amplitude modulation depth} (ranging from 0.01--10 dB). 
    We implemented this using the operation `Vibrato and tremolo' as defined in \cite{vocaltoolkit} and implemented in Parselmouth. 

\end{enumerate}

\subsection{Procedure}\label{appendix-emotion-procedure}

The main chain-construction experiment (Exp.\ 2a) assigned each participant to one of three different emotions: happiness, sadness and anger. To ensure that all participants were familiar with the emotional concept we presented contexts that has been used in previous studies on emotional prosody \citep{cowen2019, laukka2016}:

\begin{quote}
\textbf{Anger:}
Please think of a situation where you experienced a demeaning offense against you and yours. For example, somebody behaves rudely toward you and hinders you from achieving a valued goal. The situation is unexpected and unpleasant, but you have the power to retaliate.
\end{quote}

\begin{quote}
\textbf{Happiness:} 
Please think of a situation where you made reasonable progress toward the realization of a goal. For example, you have succeeded in achieving a valued goal. Your success may be due to your own actions, or somebody else’s, but the situation is pleasant and you feel active and in control.
\end{quote}

\begin{quote}
\textbf{Sadness:} 
Please think of a situation where you experienced an irrevocable loss. For example, you lose someone or something very valuable to you, and you have no way of getting back that what you want.
\end{quote}


Each participant was randomly assigned to a different emotion (happy, sad, angry). After the headphone-screening task and a short demographic questionnaire, they took a practice trial to familiarize themselves with the slider. They then completed two within-participant chains corresponding to three alterations of each of the seven dimensions, alternating between both chains until both were complete. Each chain was initialized with the feature values of the reference sentence. In each trial the participant could chose from 25 stimuli synthesized from 25 equidistant points on the slider.

In the validation experiment (Exp.\ 2c), each participant rated stimuli for the three emotion words (happiness, sadness, anger) in three corresponding randomly ordered blocks. Each block contained 49 stimuli, which came in four types:
(a) raw samples from the GSP chains, (b) samples derived by averaging the last three iterations of the GSP chains, (c) the initial unchanged sentences, (d) samples corresponding to random feature values.
Participants were presented with the same emotional contexts as the participants in the chain-construction experiment, and responded using the same four-point scale as the other experiments (`1. Not at all', `2. A little', `3. Quite a lot',  `4. Very much').


We also conducted a control experiment where we switched from within-participant chains to across-participant chains, 
reducing the number of participants by approximately half
because the original experiment proved to have more than sufficient power,
and leaving all other experiment parameters unchanged (Exp.\ 2b).
Note that reducing the number of participants should not bias the validation ratings, which only used raw samples rather than samples created by aggregating over participants.
Due to a minor implementation error, this experiment only constructed chains of length 20 rather than of length 21.

In the subsequent validation component (Exp.\ 2d), participants rated three blocks of 44 stimuli: 20 samples from the original within-participant chains, 20 samples from the new across-participant chains, 3 random samples, and one initial unchanged sentence. In all other regards this second validation was identical to the first validation.



\subsection{Supplementary results}\label{appendix-emotion-results}

\begin{figure}[h!]
    \centering
    \includegraphics{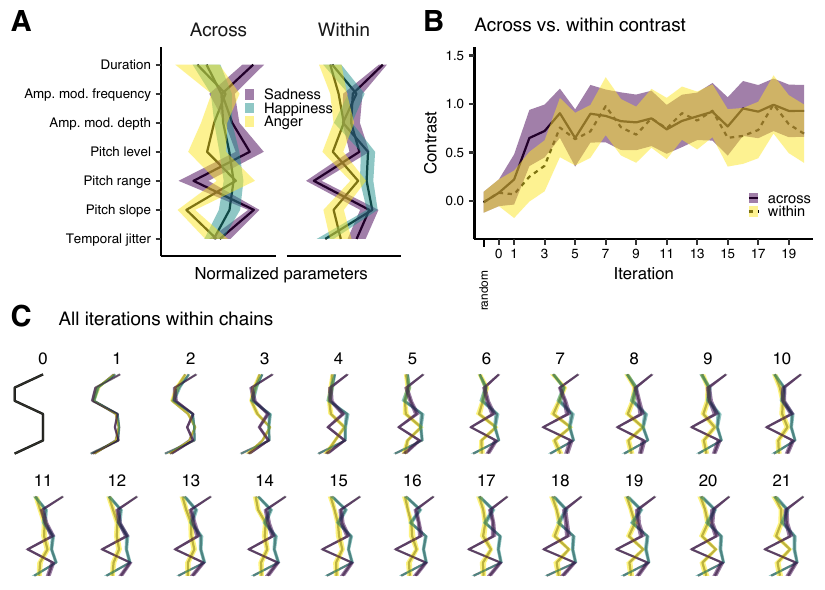}
    \caption{\textbf{A}: Average parameter settings for across- and within-participant chains in iteration 20 (Exp.\ 2a, 2b, 95\% confidence intervals over chains). \textbf{B}:  Mean validation contrast for different iterations (Exp.\ 2d, 95\% confidence intervals over participants). Contrast is defined as the difference between the rating for the target emotion and the mean rating for the non-target emotions. \textbf{C}: Average parameter settings for all iterations in within-participant chains (Exp.\ 2a).}
    \label{fig:emo-sup}
\end{figure}

Fig.\ \ref{fig:emo-sup}A shows the results of the within- and across-participant comparison (Exp.\ 2a, 2b). The resulting feature values are broadly similar between these two experiments, suggesting that memory effects did not substantively contaminate our within-participant chains. This conclusion is supported by the validation experiment, which shows similar contrast scores for both within- and across-participant chains (Fig.\ \ref{fig:emo-sup}B, Exp.\ 2d).

Fig.\ \ref{fig:emo-sup}C shows how mean feature values develop over the course of the within-participant experiment (Exp.\ 2a). Here we can see how most of the development of the feature values occurs over the first sweep of the feature vector (iterations 1--7), after which point the feature values stay broadly similar. Three audio examples from the same sentences in the final iteration of this experiment can be found at \url{https://doi.org/10.17605/OSF.IO/RZK4S} in the folder \texttt{sound-examples-prosody}
with the filenames \texttt{sad|happy|angry\_final\_sentence.wav},
alongside for reference the initial stimulus \texttt{original\_sentence.wav}.

\begin{figure}[h!]
    \centering
    \includegraphics{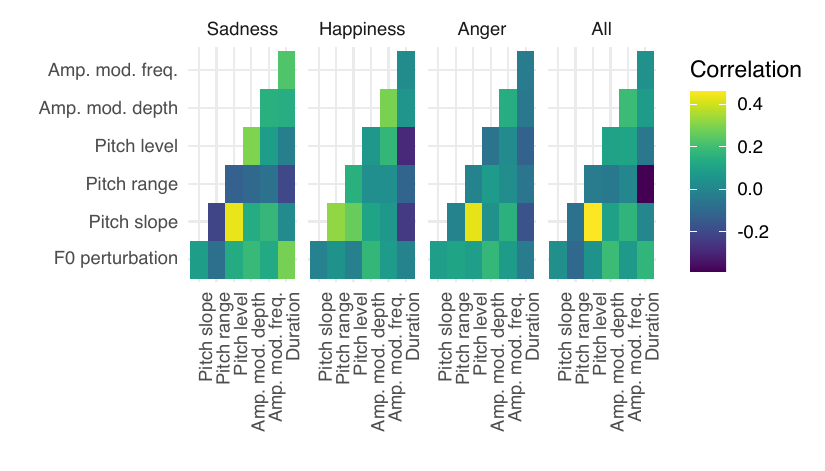}
    \caption{Pearson correlations between parameters in all three emotions (Exp.\ 2a).}
    \label{fig:emo-cor}
\end{figure}

GSP also allows us to investigate higher-order structure in perceptual representations. As an illustrative analysis, Fig.\ \ref{fig:emo-cor} plots pairwise correlations for different features in the generated samples (Exp.\ 2a). For example, we see that duration and F0 perturbation were significantly correlated for sadness (\textit{r} = .28) but not for the other emotions (anger: \textit{r} = $-$.03, happiness: \textit{r} = .00); in contrast we see that pitch level and pitch slope were positively correlated for all three emotions.
These kinds of higher-order analyses provide a more expressive perspective on prosody features than previous research, which mainly focuses on the independent contributions of single features rather than interactions between features. We intend to explore these kinds of interactions more in future research.

\section{Musical chords}\label{appendix-chords}


\begin{figure}[h]
  \centering
  \includegraphics[width=1.0\linewidth]{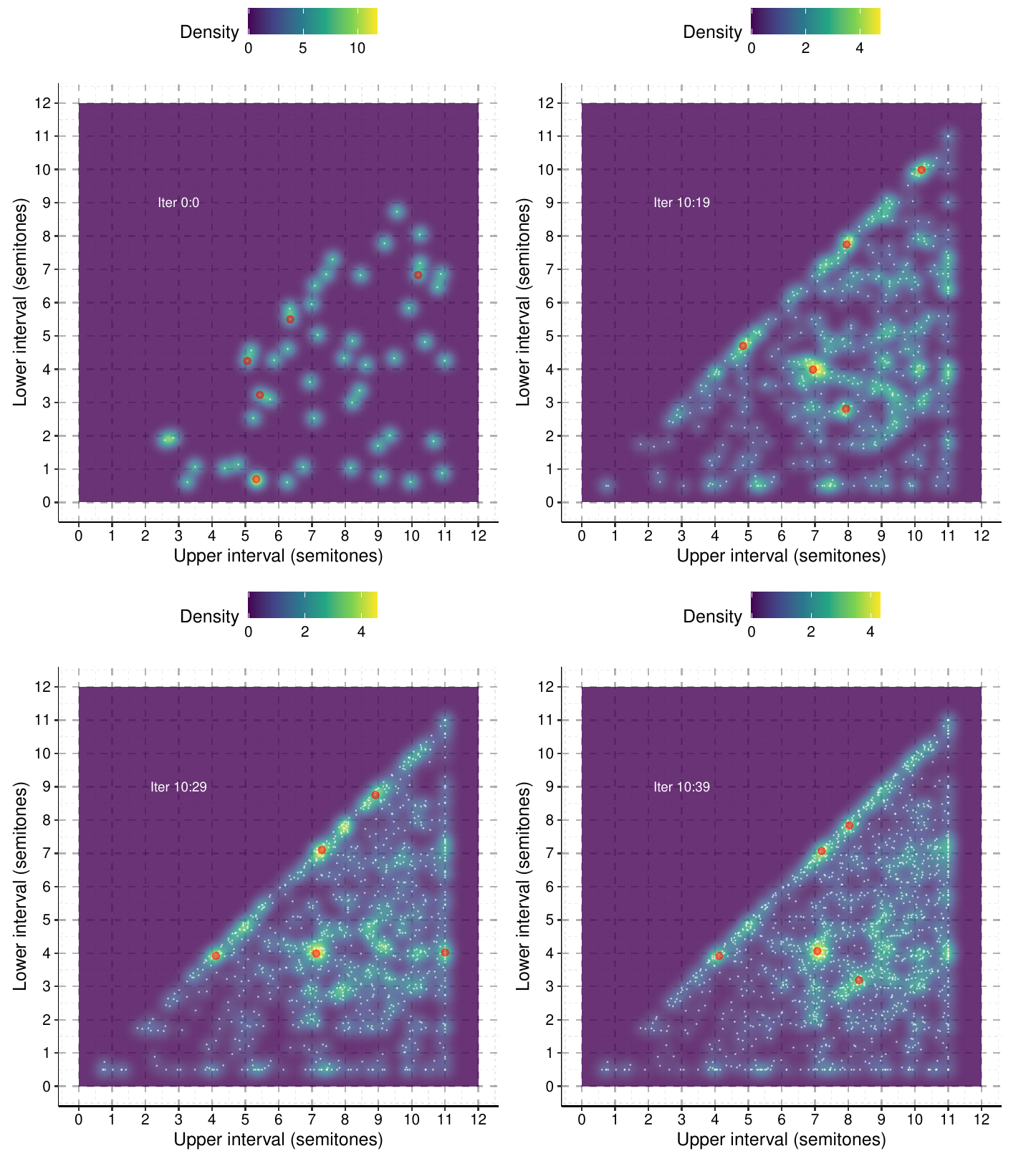}
  \caption{Kernel density estimates generating the four sets of KDE modes considered in the validation experiment for musical triads (Exp.\ 3b). The top five modes are indicated in red. The density values are computed relative to a uniform distribution.}
  \label{fig:validationKDE}
\end{figure}

\begin{figure}[h]
  \centering
  \includegraphics[width=0.6\linewidth]{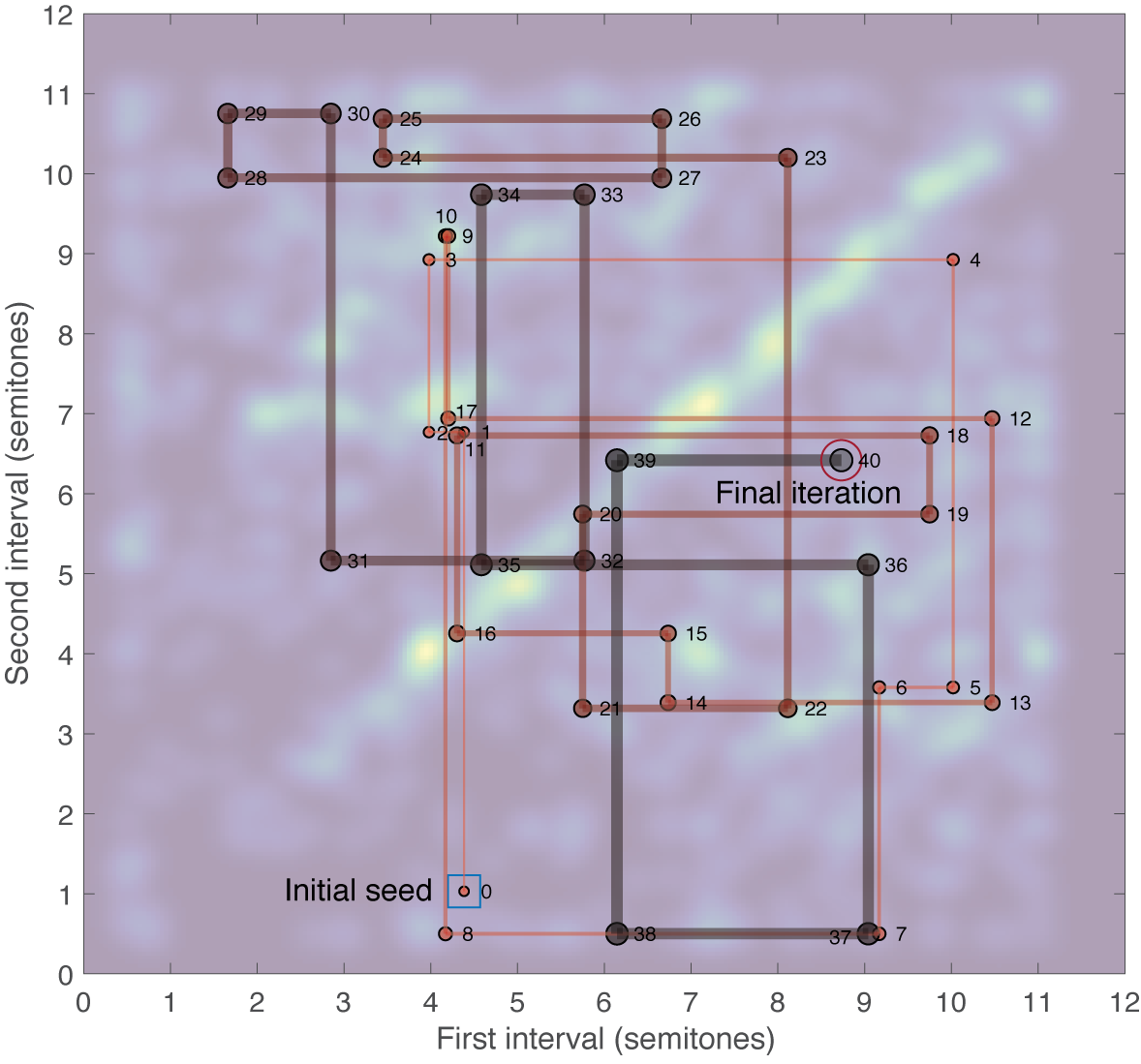}
  \caption{An example trajectory of a GSP chain over chords, layered on top of a KDE of aggregated data from iterations 10--39 (bandwidth = 0.175 semitones, Exp.\ 3a). Similar dynamics were apparent in all other chains.}
  \label{fig:trajectories}
\end{figure}

\begin{figure}[h]
  \centering
  \includegraphics[width=0.6\linewidth]{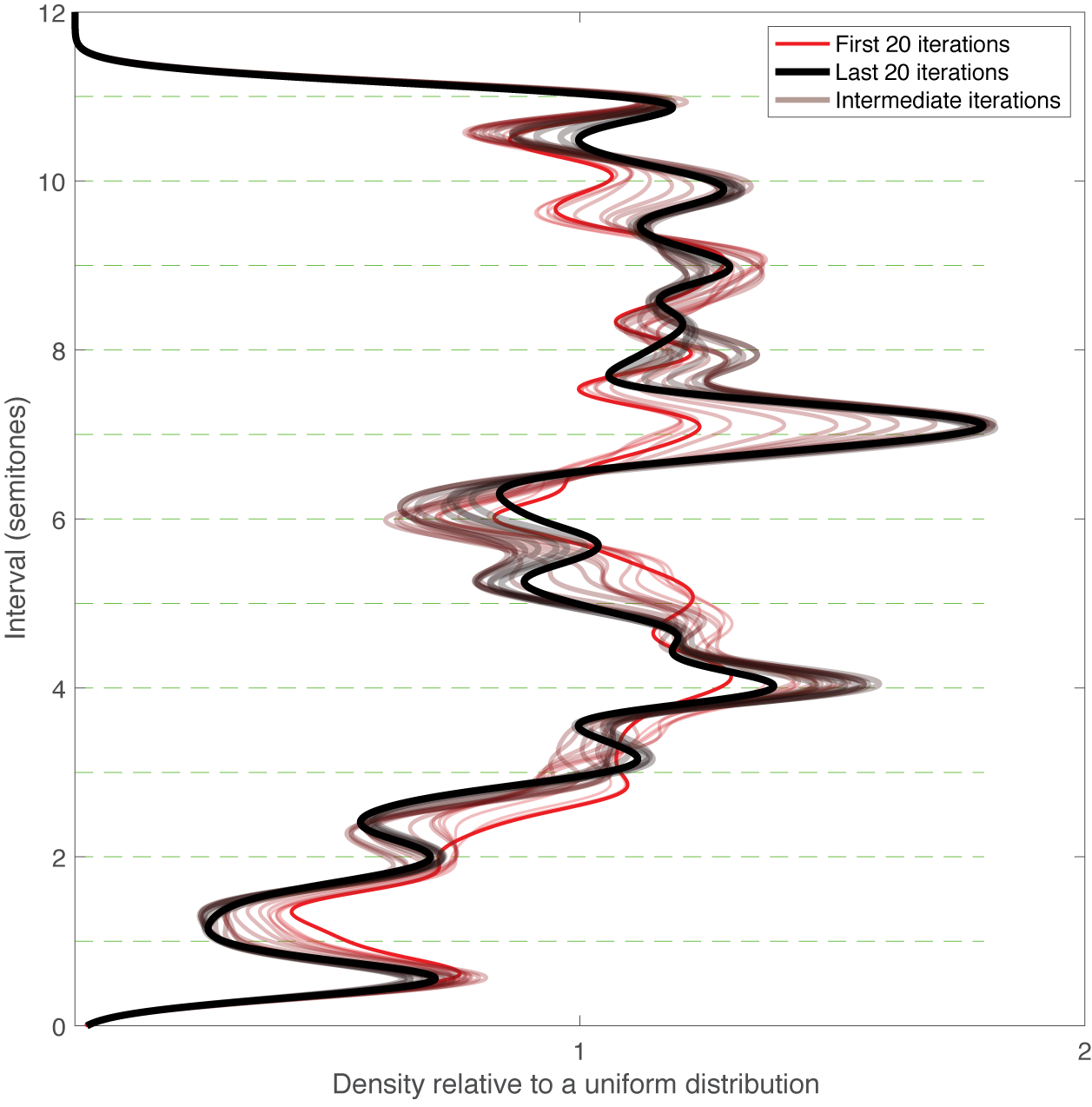}
  \caption{Combined marginal distributions for the two intervals, using a sliding window of length 20 (Exp.\ 3a).}
  \label{fig:dynamicMarginal}
\end{figure}

\subsection{Supplementary methods}\label{appendix-chords-methods}

This study applied GSP to the perceived pleasantness of musical chords. Each of these chords comprises three tones, and is hence termed a \textit{triad}. We represented each triad as a pair of numbers, following the `pitch chord type' representation of \citep{HarrisonHREP}, which represents each chord tone as a pitch interval in semitones from the bass (i.e., lowest) tone. This representation captures the sense in which human pitch perception is relative (i.e., pitches are heard relative to their recent auditory context) and logarithmic (i.e., perceived pitch distance is approximately proportional to the difference in the logarithm of the frequencies) \citep{Stainsby2009}. Integer values in this representation correspond to the standard 12-tone equal-tempered tuning system of Western music. 

We generated chords using Tone.js, a Javascript library for synthesizing sounds in the client's browser.\footnote{\url{https://tonejs.github.io/}}
Each triad was synthesized as three simultaneous complex tones comprising 10 harmonics with amplitudes scaled by 12 dB/octave. These complex tones were presented with an ADSR envelope comprising a linear attack portion of 200 ms and a maximum amplitude of 1.0, an exponential decay portion lasting 100 ms taking the amplitude to 0.8, and a final exponential decay release portion lasting 1 s.  The pitch of the bass tone was sampled uniformly and continuously in the logarithmic range G3--F4 (i.e., 196--349 Hz).
The other two tones were specified by two continuous intervals in the range [0.5, 11], with the limits chosen such that the unison (0) and octave (12) were excluded, to prevent duplicating the pitch class of the bass tone.\footnote{
Two tones are said to share the same pitch class if they are separated by an integer multiple of 12 semitones (an octave).}
We did however allow the two non-bass tones to overlap.



In each trial of the main experiment (Exp.\ 3a), participants were presented with the following prompt: `Adjust the slider to match the following word as well as possible: pleasant'. Releasing the slider prompted a chord to be played whose pitch intervals reflected the current position of the slider.
Before beginning the main experiment, participants completed three training examples to familiarize themselves with the procedure. 

Through this experiment we constructed 50 across-participant chains comprising 40 iterations. The starting seed for each chain was created by randomly sampling both intervals from a uniform distribution in the range [0.5, 11].
Each participant contributed up to 20 trials to these chains.



The validation experiment (Exp.\ 3b) collected ratings for various raw samples and KDE modes, extracted from various iterations of the 50 chains constructed in the GSP experiment. The raw samples were represented by 16 experimental conditions corresponding to the iterations 0, 1, ..., 9 and 14, 19, ..., 39, with each condition being represented by 50 stimuli (1 stimulus from each chain). In addition to the raw sample conditions, there were four KDE conditions, with each KDE condition being represented by 5 modes, and the modes being extracted by aggregating data from iterations 0, 10 to 19, 10 to 29 and 10 to 39 (Fig.\ \ref{fig:validationKDE}); here iteration 0 corresponds to the random seed. We used Gaussian kernels of width 0.175, and extracted modes using the kernel-based clustering algorithm of \citep{rodriguez2014clustering} as implemented in the \texttt{ADPclust} R package \citep{wang2017fast} with the number of clusters set \textit{a priori} to 20, and keeping the five resulting centroids with the highest kernel density. 
In total, this resulted in 20 experimental conditions comprising 820 stimuli in total. 

In each trial of the validation experiment, the participant was assigned to a randomly chosen stimulus from one of the conditions, and was asked to rate how pleasant that stimulus was on a four-level scale: `Not at all', `A little', `Quite a lot' and `Very much'. Overall we collected 662 ratings for each experimental condition, 
with each participant contributing up to 80 ratings. 

We should note that while in Gibbs sampling it is customary to consider samples in jumps of full coordinate sweeps, here we decided to aggregate data continuously, given the inherent symmetry between the two intervals, so as to improve the quality of the estimated modes. We further exploited that symmetry by folding the data along the $x = y$ line, since reordering a pair of intervals does not alter the generated chord.
Fig.\ \ref{fig:trajectories} shows the raw data, where the $x = y$ symmetry is clearly apparent, and Fig.\ \ref{fig:validationKDE} shows the folded distribution after re-ordering the two intervals.

\subsection{Supplementary results}\label{appendix-chords-results}

In the main paper, we mostly discussed the structure and validation of features and raw samples aggregated across various chains and iterations. Fig.\ \ref{fig:trajectories} complements this perspective by presenting the trajectory of a typical chain (Exp.\ 3a). It is clear that the dynamics are far from an optimization regime, where one would expect to see small and converging updates toward some local optimum (e.g., \citep{jacoby2017integer}). Instead, the trajectories illustrate the sampling regime of GSP, characterized by big leaps and lack of convergence, scanning the various regions of the space.

Fig.\ \ref{fig:dynamicMarginal} shows the behavior of the combined marginal distributions for the two intervals, computed over a sliding window of length 20 with Gaussian kernels (bandwidth = 0.175 semitones, Exp.\ 3a). 
We see that two strong modes emerge at the perfect fifth (7) and the major third (4),
alongside other peaks at integers and dips at the semitone (1) and tritone (6), reflecting the standard Western tonal hierarchy \citep{krumhansl1982tracing}.

Audio samples of the top 15 KDE modes extracted from iterations 0 (random) and 10--39 can be found at \url{https://doi.org/10.17605/OSF.IO/RZK4S} in the folder \texttt{sound-examples-musical-triads}, with the modes arranged in descending order of density (\texttt{random\_seed\_top\_modes\_iter\_0\_0.wav} and \texttt{top\_modes\_iter\_10\_39.wav}).

\section{Faces}\label{appendix-faces}


\subsection{Supplementary methods}\label{appendix-faces-methods}

This study used the `StyleGAN' model of \cite{karras2019style, stylegan2} pretrained on the FFHQ dataset of faces from Flickr \cite{karras2019style}.
This model is a generative adversarial network, comprising a latent vector $\mathbf{z}$ sampled from a probability distribution $p(\mathbf{z})$, an input layer that takes a constant input $\mathbf{y}_0$, and the other layers $\mathbf{y}_i$ taking the previous layer and a non-linear function of $\mathbf{z}$ as an input:

\begin{equation}
\mathbf{y}_i = G_i(\mathbf{y}_{i-1}, \mathbf{w}), \quad
\mathbf{w} = M(\mathbf{z}),
\end{equation}

where $M$ is an 8-layer multilayer perceptron and the output layer $\mathbf{y}_L$ corresponds to an RGB image. 

The study depended on participants interactively manipulating principal components of the $\mathbf{w}$ vector using a slider. We achieved this by creating an API that took as input a random seed for the latent vector $\mathbf{z}$, a vector of principal component values for $\mathbf{w}$, and the index of the principal component to be manipulated by the slider. The API then returned a video where the active principal component was incrementally modified through a specified number of standard deviations about the mean, with this API building on code released in \cite{ganspace}. The resulting video was then streamed to the participant's local computer, with the slider selecting between different frames of the video. We hosted the API on an AWS EC2 instance fitted with an NVIDIA K80 GPU.

An important technical issue concerned ensuring that participants didn't have to wait for the relatively slow stimulus generation process. We therefore generated stimuli asynchronously in advance of a given experimental trial, with participants being randomly assigned to the pool of currently available stimuli for each trial. Aggregating multiple responses per step of the GSP process helped in this regard, meaning that a higher throughput of participants could be sustained for a given rate of stimulus production.

The main experiment (Exp.\ 4a) evaluated six adjectives which we thought could elicit meaningful perceptual associations: `attractive', `fun', `intelligent', `serious', `trustworthy', and `youthful', with these choices informed by prior literature (e.g., \citep{brinkman2017visualising}).
Three across-participant chains were constructed for each of these adjectives, each of length 50 plus the initial random state, resulting in a total of 18 chains. Each step in the chain received five responses from five different participants, which were then aggregated using the arithmetic mean.

Participants were recruited from AMT as before with the stipulation that they be resident in the US. All participants were pre-screened with the color vocabulary task used previously for the color experiment. After completing a short demographic questionnaire, they took six practice trials to familiarize themselves with the task, then proceeded to the main experiment, where they completed up to 18 trials (one from each chain). 


The validation experiment (Exp.\ 4b) recruited participants in the same manner, and had the participants rate all generated samples from iterations 1--10, 20, 30, 40, and 50. Each participant contributed 80 ratings, under the constraint that they never rated the same sample twice, and with participants being assigned to stimuli such that the number of ratings accumulated equally across stimuli. Data collection was continued until all samples had been rated at least 50 times.


We additionally conducted several follow-up GSP experiments to explore the paradigm further, described below and in Table \ref{experiments-table}:

\begin{enumerate}
    \item We tested an alternative aggregation approach, where we summarized the five responses for each item with a KDE (Gaussian kernel, standard deviation of 0.5 in units of PCA standard deviations), and took the mode of the resulting distribution 
    (Exp.\ 4c, Fig.\ \ref{fig:supplementary-face-validation}).
    \item We tested a small number of alternative methods for constructing a basis for the stimulus space (Exp.\ 4e, Fig.\ \ref{fig:supplementary-face-basis}). In addition to the original PCA, we tested sparse PCA using a sparsity parameter of 1.0 (see the alpha parameter of \texttt{SparsePCA} from the \texttt{scikit-learn} package) and independent component analysis (ICA). 
    We also tested the effect of retaining dimensions 71--80 instead of dimensions 1--10 of the PCA solution.
    In this experiment we only used the adjective `attractive', and reduced the chain length to 30 iterations. For comparability with the original results of Exp.\ 4a, all chains were initialized to the same random seeds as in the original experiment.
    \item We reran the original experiment but with the StyleGAN model pretrained on a dataset of faces from WikiArt (\url{https://www.wikiart.org}; \url{https://github.com/ak9250/stylegan-art}), to illustrate the dataset-dependence of the results (Exp.\ 4h, Fig.\ \ref{fig:art-faces}).
    \item We reran the original experiment using KDE modes and relaxing participant recruitment to accept both US and non-US participants (Exp.\ 4i, Fig.\ \ref{fig:supplementary-face-validation}). The resulting participant group was dominated by Indian (c. 50\%) participants but also included a high proportion of US participants (c. 40\%).
    \item We reran the original experiment but asking participants to adjust the slider to `find the person that you would most like to date', assigning self-reported male and female participants to separate chains so that we could perform a group-difference analysis (Exp.\ 4j, Fig.\ \ref{fig:faces-dating}).
\end{enumerate}

We additionally ran several rating experiments to complement these GSP experiments (Table \ref{validation-experiments}). 
Exp.\ 4d collected ratings for the KDE mode experiment (Exp.\ 4c) and the global participant group experiment (Exp.\ 4i), as well as collecting ratings for the original experiment (Exp.\ 4a), with otherwise the same design as the original validation experiment (Exp.\ 4b), including the US-only criterion
(Fig.\ \ref{fig:supplementary-face-validation}). 
Exp.\ 4f used the same approach to collect ratings for the basis experiment (Exp.\ 4e, Fig.\ \ref{fig:supplementary-face-basis}). 

Exp.\ 4g used a similar design to investigate biases at different stages of the modeling pipeline
(Fig. \ref{fig:bias-gender}, \ref{fig:bias-gender-age}, \ref{fig:bias-pipeline-features-gender}, \ref{fig:bias-gsp-features}). 
Stimuli were sourced from three stages: (a) random samples from the StyleGAN's FFHQ training dataset (\textit{N} = 300); (b) random samples from the StyleGAN model (\textit{N} = 300); random samples from the StyleGAN model, but only allowing the top 10 principal components to vary (\textit{N} = 300); (c) samples from iterations 0, 10, 20, 30, 40, and 50 of the GSP processes from Exp.\ 4a (\textit{N} = 108). Instead of asking participants to rate how well the images matched the GSP adjectives, we instead asked participants to answer questions from the following list:

\begin{enumerate}
    \item What is the gender of the person in the image?
    \item Is the person in the image of white ethnicity?
    \item Is the person in the image smiling?
    \item Is the person in the image wearing a hat?
    \item Is the person in the image wearing formal clothes?
    \item Is the person in the image wearing glasses?
\end{enumerate}

In each case, the participant was presented with three options:
``Male''/``Female''/``Other'' in the case of gender, and 
``Yes''/``No''/``Don't know'' in the other cases.
We also asked participants to estimate the age in years of the person depicted in the image.

We had two kinds of motivations for choosing these particular evaluations.
We chose gender, age, and ethnicity because these are two criteria 
according to which many people experience bias in the real world,
and we wanted to understand how these variables were 
treated by the modeling pipeline.
We chose the other four evaluations because they are examples of 
easily quantified features that seem likely to influence judgments made about the person.
Of course, it should be acknowledged that some of these variables are impossible to 
determine definitively from an image; for example, it is a substantial 
simplification to treat gender and ethnicity in a categorical way.
However, we anticipated that this simplification would be necessary to make the task
understandable to the participants, and that the resulting data
would nonetheless be informative about the kinds of biases present in the 
modeling pipeline.


\subsection{Supplementary results}\label{appendix-faces-results}

\begin{figure}
  \centering
  \includegraphics[width=1.0\linewidth]{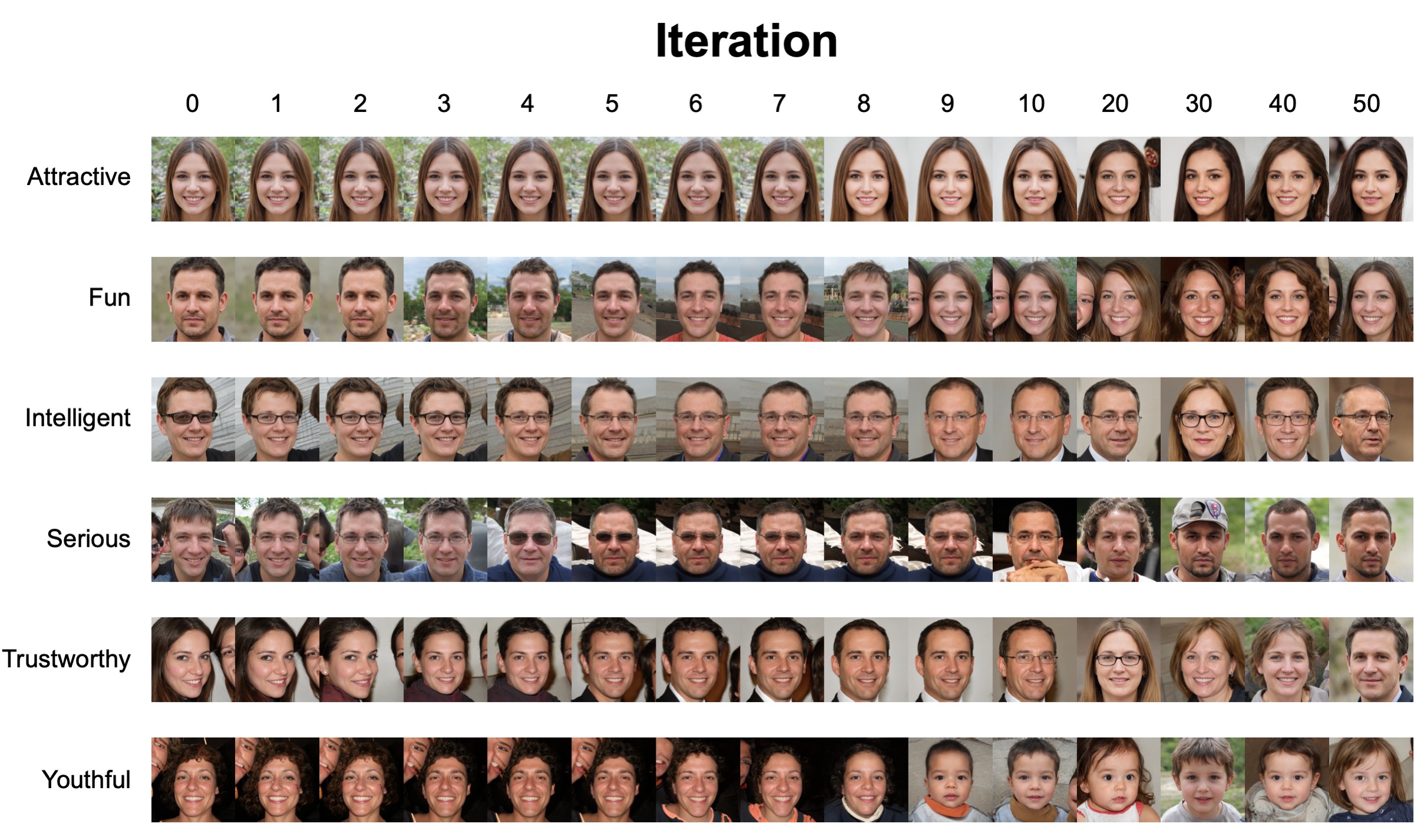}
  \caption{Raw samples from six GSP chains in Exp.\ 4a (US-only participants, mean aggregation).}
  \label{fig:faces-us-mean}
\end{figure}

\begin{figure}[h]
  \centering
  \includegraphics[width=1.0\linewidth]{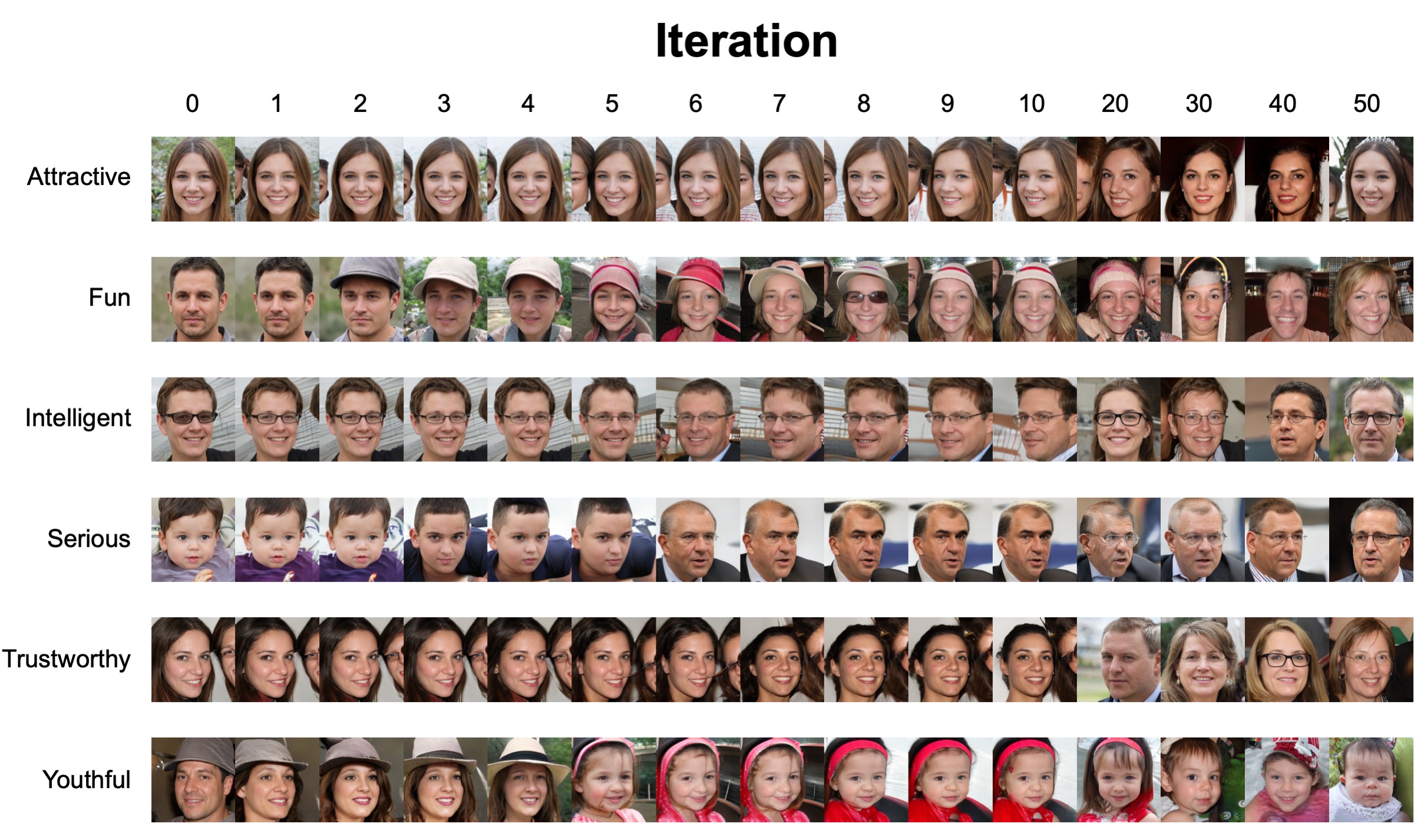}
  \caption{Raw samples from six GSP chains in Exp.\ 4c (US-only participants, KDE mode aggregation).}
  \label{fig:faces-us-mode}
\end{figure}

\begin{figure}[h]
  \centering
  \includegraphics[width=1.0\linewidth]{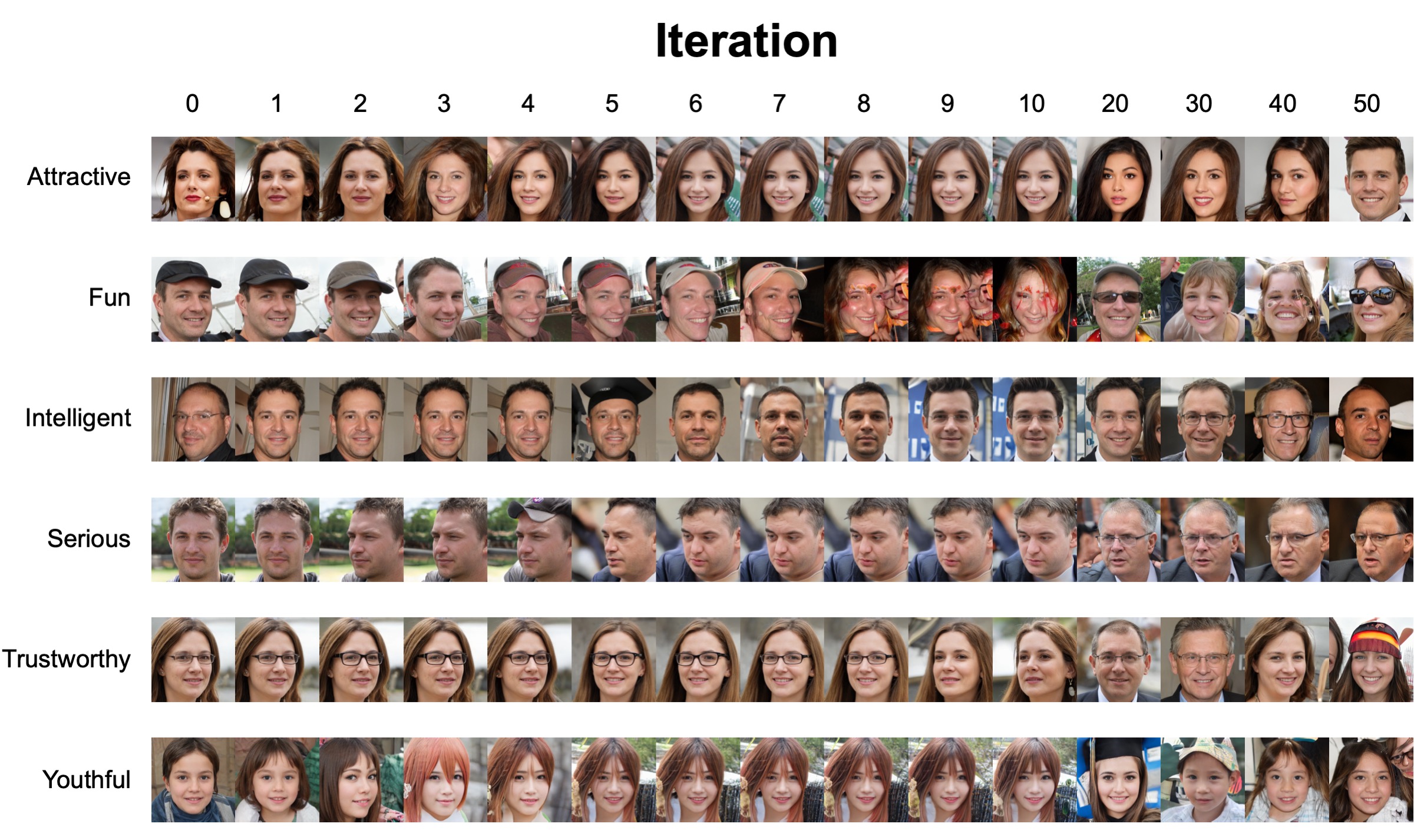}
  \caption{Raw samples from six GSP chains in Exp.\ 4i (global participant group, KDE mode aggregation).}
  \label{fig:faces-non-us-mode}
\end{figure}

\textbf{Raw samples from the main experiment.} Example raw samples and validation results from the main experiments (Exp.\ 4a, 4b) are shown in Fig.\ 4 of the main paper. Fig.\ \ref{fig:faces-us-mean} illustrates the raw samples in more detail, displaying iterations 0--10, 20, 30, 40, 50 from one chain for each target word. It is clear from both the validation results and the raw samples that the chains make clear progress towards the target category already by the end of the first sweep (10 iterations), and sometimes the resemblance to the target category does not improve noticeably after this point. However, this does not mean that the process converges to a static image after this point: instead, there is a moderate amount of variety in the subsequent faces (see also Fig.\ \ref{fig:faces-us-mode} and \ref{fig:faces-non-us-mode}). The process is therefore still somewhat in the stochastic sampling regime rather than the deterministic optimization regime.

\begin{figure}
  \centering
  \includegraphics[width=1.0\linewidth]{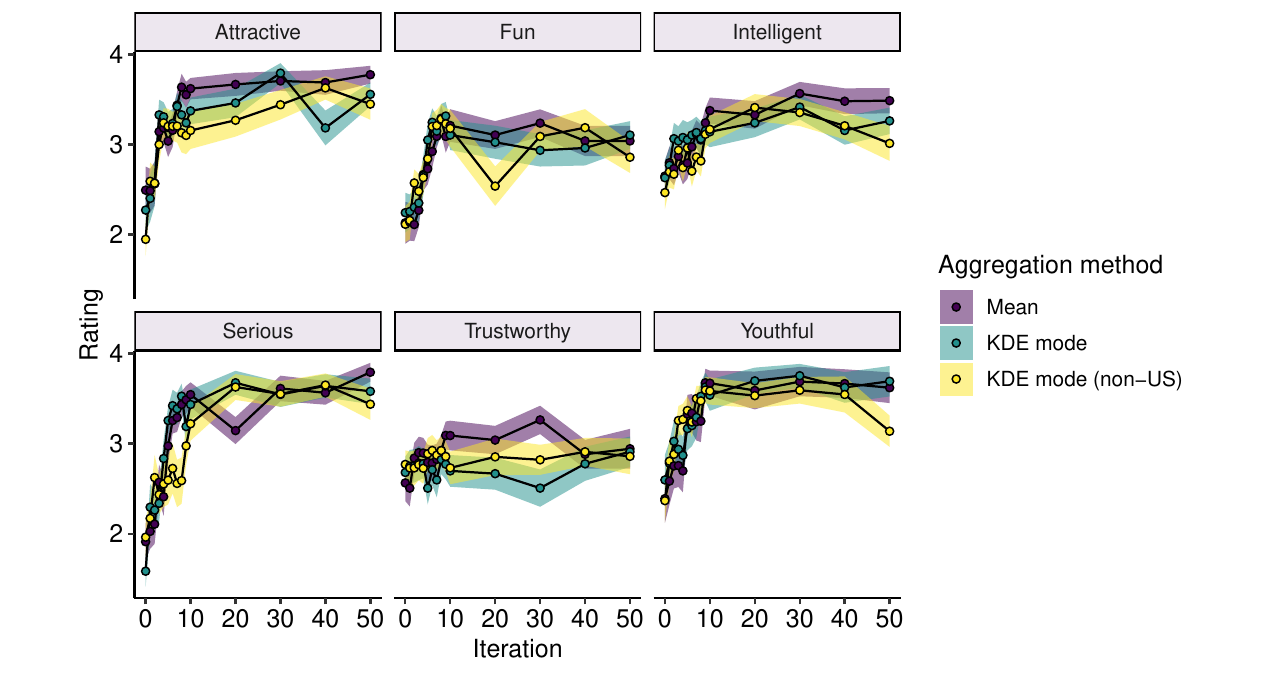}
  \caption{Validation results for Exp.\ 4a, 4c, and 4i, as produced in Exp.\ 4d. The shaded regions correspond to 95\% confidence intervals over participants.}
  \label{fig:supplementary-face-validation}
\end{figure}

\textbf{Validation results for follow-up experiments. }Fig.\ \ref{fig:supplementary-face-validation} plots validation results for Exp.\ 4a (mean aggregation, US-only participants), Exp.\ 4c (aggregation with KDE modes, US-only participants), and Exp.\ 4i (aggregation with KDE modes, global participants), as collected in Exp.\ 4d. The broad trends in the ratings are replicated across the three experiments: typically almost all of the improvement comes in iterations 1--10, with ratings staying mostly stable after this point. All three experiments struggle to capture trustworthiness, which is clearly a particularly subjective judgment to make. Interestingly, there is no evidence that KDE peak-picking outperforms the arithmetic mean as an aggregation technique. Inspecting the raw data and the density estimates, this does not seem to be a consequence of poorly chosen kernel width or artifacts in the density estimation process. Instead, it seems that the participants' conditional distributions could typically be approximated well by a unimodal distribution, and hence averaging was a sensible aggregation method.

\textbf{Hints at cross-cultural differences.} Raw samples of six chains from Exp.\ 4c (aggregation with KDE modes, US-only participants), and Exp.\ 4i (aggregation with KDE modes, global participants) are displayed in Fig.\ \ref{fig:faces-us-mode} and \ref{fig:faces-non-us-mode}. It is important not to read too much into these raw samples, as they ultimately come from stochastic distributions and will vary over repeated runs. However, we did notice some suggestive differences between the final samples of the US chains and those of the global chains. Most salient was the fact that all US chains for `intelligent' finished with a Caucasian man, whereas the three final states of the global chains included both a woman and a non-Caucasian man. We also noticed that the global chains were the only ones to include a man as the final `attractive' sample. While some of this variation will be due to chance, the remaining variation will presumably reflect different stereotypes held by the different participant groups. It would be interesting to explore these different stereotypes in more systematic ways.

\begin{figure}[h]
  \centering
  \includegraphics[width=0.8\linewidth]{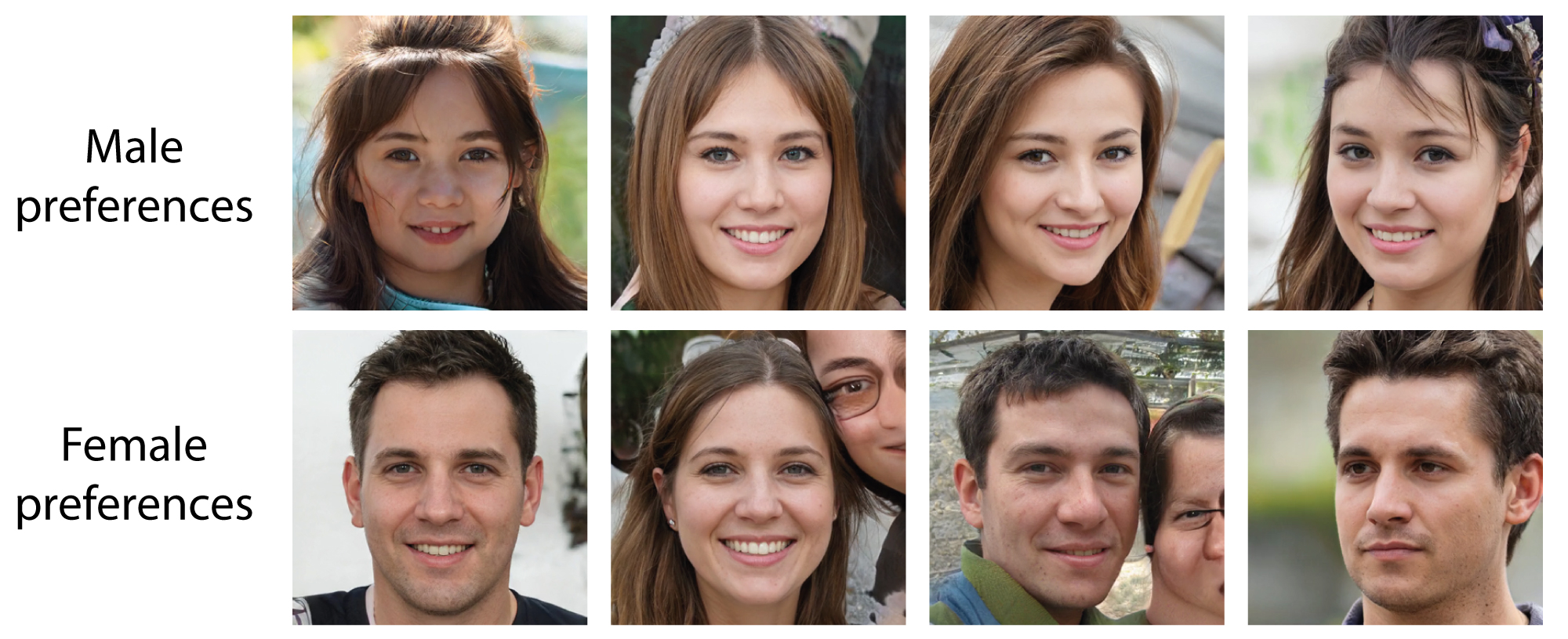}
  \caption{Final samples from the first four male and female chains in the dating preferences experiment (Exp.\ 4j).}
  \label{fig:faces-dating}
\end{figure}

\textbf{Gender differences.} Exp.\ 4j provides a second proof of concept for this kind of group-difference approach (Fig.\ \ref{fig:faces-dating}).
Here participants were split by self-reported gender, and instructed to optimize the slider for a person that they would most like to date.
As one might expect, the samples reflect a predominant (but not universal) preference for members of the opposite gender. 
This in itself may be a trivial result, but it is easy to intuit how one could extrapolate this approach to much more complex and interesting group-difference studies, for example those involving different cross-cultural populations.

\begin{figure}
  \centering
  \includegraphics[width=0.7\linewidth]{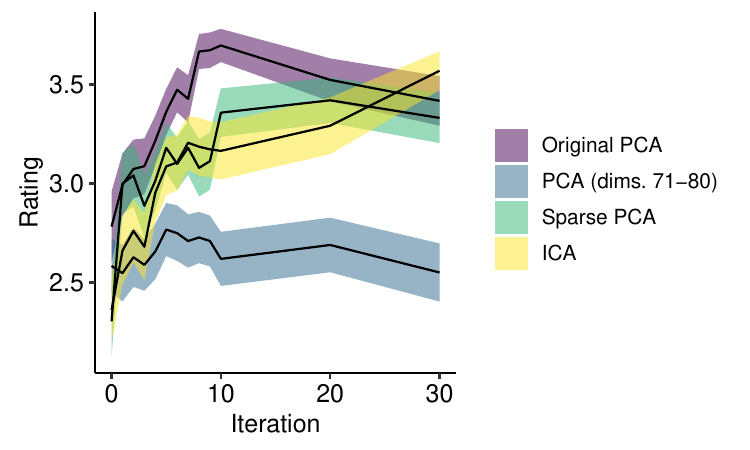}
  \caption{Validation results for Exp.\ 4e (exploring different basis construction methods), as collected in Exp.\ 4f. The shaded regions correspond to 95\% confidence intervals over participants.}
  \label{fig:supplementary-face-basis}
\end{figure}

\textbf{Basis construction methods.} Fig.\ \ref{fig:supplementary-face-basis} plots validation results for Exp.\ 4e (exploring different basis construction methods), as collected in Exp.\ 4f. The results suggest an early advantage for the original PCA technique; however, the discrepancy with sparse PCA and ICA is small, and seems to disappear after more iterations. As would be expected, the version of PCA with components 71--80 performs poorly; in practice, these components contribute very little perceptually speaking (see also \citep{ganspace}). On this basis, there is little evidence to dismiss any one of PCA, sparse PCA, or ICA. Future work should also consider other recently proposed approaches for parameterizing the generative model, for example \citep{voynov2020,shen2020}.

\begin{figure}
  \centering
  \includegraphics[width=0.7\linewidth]{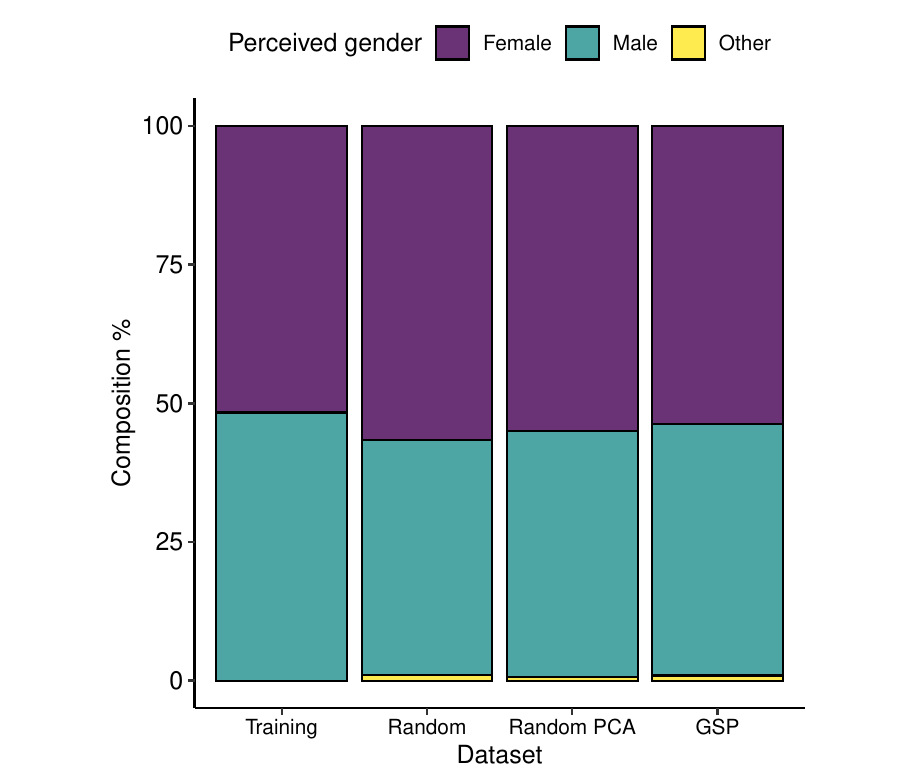}
  \caption{Perceived gender for faces from different stages of the modeling pipeline, as collected in Exp.\ 4g.}
  \label{fig:bias-gender}
\end{figure}

\begin{figure}
  \centering
  \includegraphics[width=0.7\linewidth]{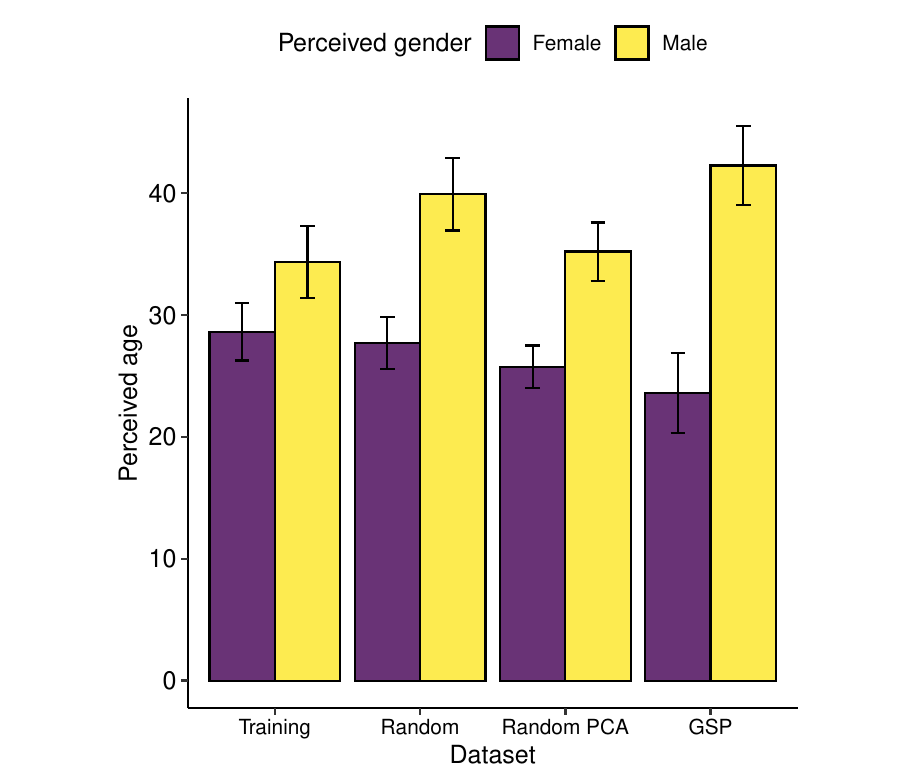}
  \caption{Perceived age split by gender for faces from different stages of the modeling pipeline, as collected in Exp.\ 4g. The error bars denote 95\% confidence intervals bootstrapped over images.}
  \label{fig:bias-gender-age}
\end{figure}

\textbf{Bias analyses.} Fig.\ \ref{fig:bias-gender} plots perceived gender in the different datasets evaluated in Exp.\ 4g.
We see that the gender balance is fairly equal between men and women,
with perhaps slightly more women than men as the pipeline progresses.
Fig.\ \ref{fig:bias-gender-age} plots perceived age as a function of perceived gender in the same datasets.
Looking first at the training dataset, we see that the mean age is close to 30 years, with the male faces tending to be perceived as somewhat older than 30, and the female faces being perceived as slightly younger than 30.
This association between age and gender is amplified to a certain amount through the modeling pipeline, even before the PCA process;
it seems as if the model is capturing this association and stereotyping it to a certain degree.
This relationship has interesting implications for the GSP samples;
if female samples tend to be subjectively younger than male samples, and if younger faces tend to be perceived as more attractive, then GSP samples for `attractive' will be biased towards women, even if the participants do not possess any systematic bias for women over men. 
Likewise, if older faces tend to be perceived as more intelligent, then this relationship between age and gender would be expected to induce a bias in the GSP samples for `intelligent' towards male faces.

\begin{figure}
  \centering
  \includegraphics[width=1.0\linewidth]{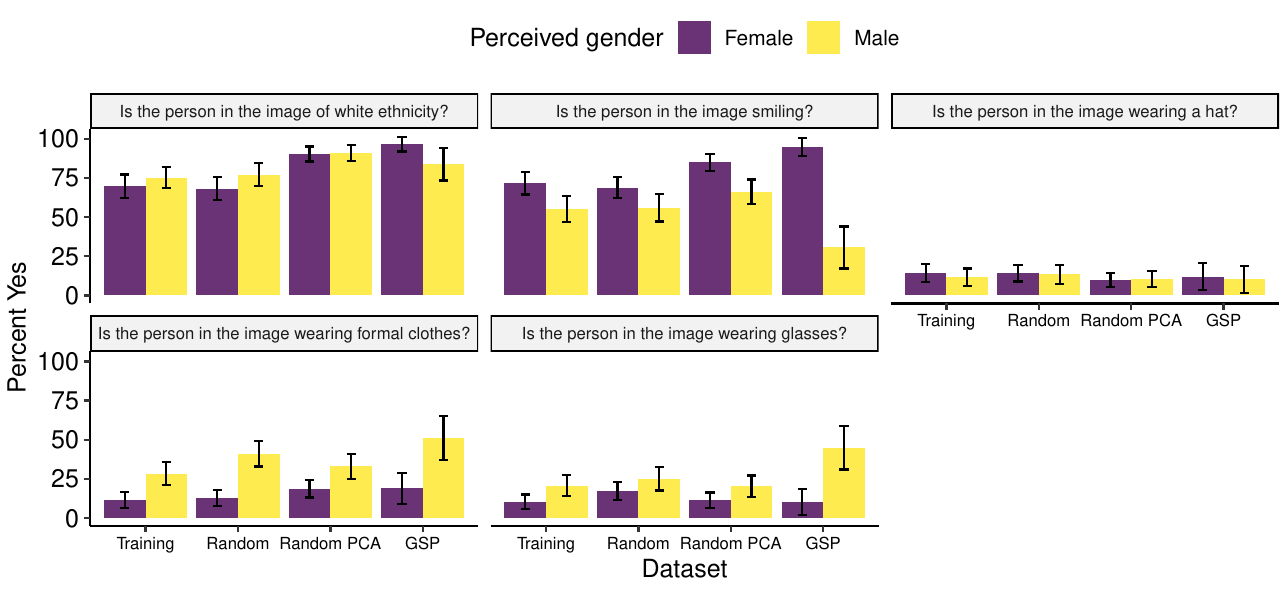}
  \caption{Evaluations of ethnicity, smiling, hats, formal clothes, and glasses, for faces from different stages of the modeling pipeline, split by gender (Exp.\ 4g). The error bars denote 95\% confidence intervals bootstrapped over images.}
  \label{fig:bias-pipeline-features-gender}
\end{figure}

\begin{figure}
  \centering
  \includegraphics[width=1.0\linewidth]{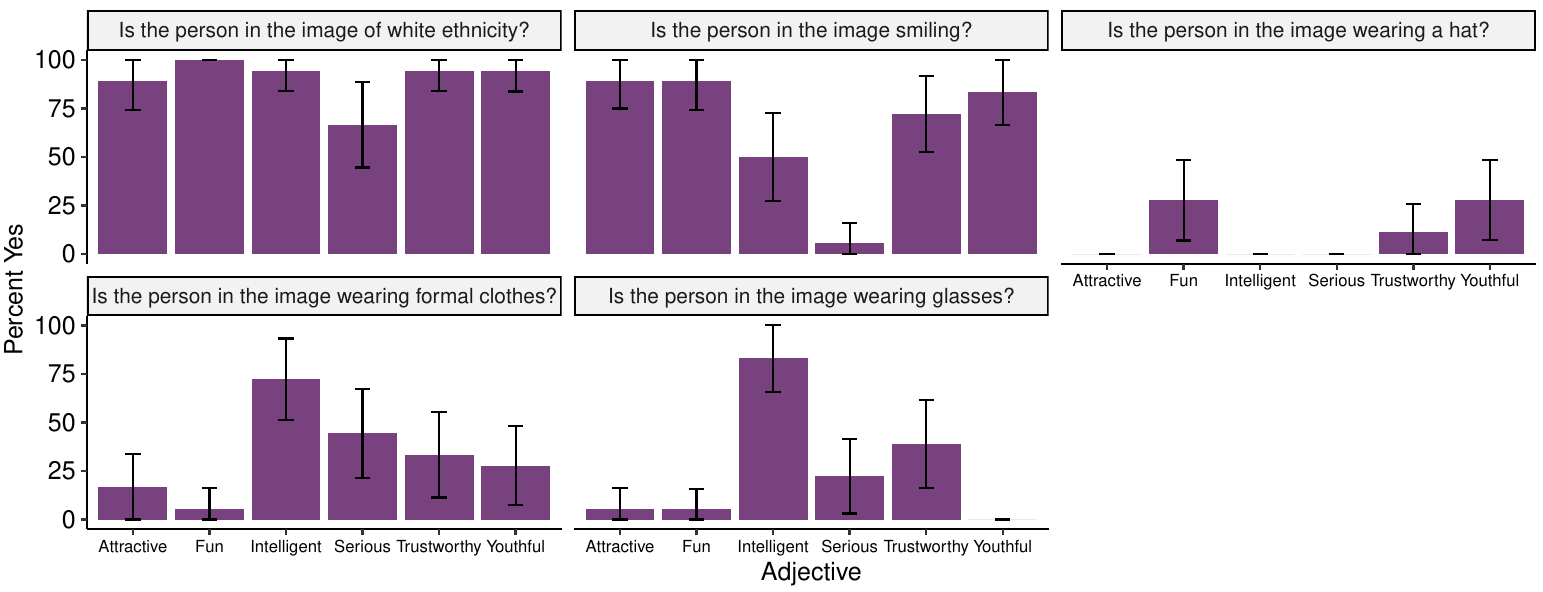}
  \caption{Evaluations of ethnicity, smiling, hats, formal clothes, and glasses, for GSP samples evaluated in Exp.\ 4g. The error bars denote 95\% confidence intervals bootstrapped over images.}
  \label{fig:bias-gsp-features}
\end{figure}

There are many other similar biases that one could anticipate affecting the GSP process.
To illustrate some of these potential biases, Fig.\ \ref{fig:bias-pipeline-features-gender} plots judgments for ethnicity, smiling, hats, formal clothes, and glasses wearing for the four datasets in Exp.\ 4g, split by gender.
We see for example that men are much more likely than women to be portrayed in formal clothes, potentially a further reason why `intelligent' GSP samples tend to favor men. 
Similarly, men are more likely to be portrayed in glasses, another potential contributor to perceived intelligence. Conversely, women are more likely than men to be smiling, potentially supporting a female bias in the `attractive' and `fun' GSP samples. 
These hypotheses are consistent with Fig.\ \ref{fig:bias-gsp-features}, which shows that perceived intelligence is indeed associated with wearing formal clothes and glasses, and that perceived attractiveness and fun are both associated with smiling.
These examples illustrate the complex network of biases that can be inherited from a generative model such as StyleGAN, and highlight the importance of developing more balanced training datasets for future cognitive work in this area.

\begin{figure}
  \centering
  \includegraphics[width=1.0\linewidth]{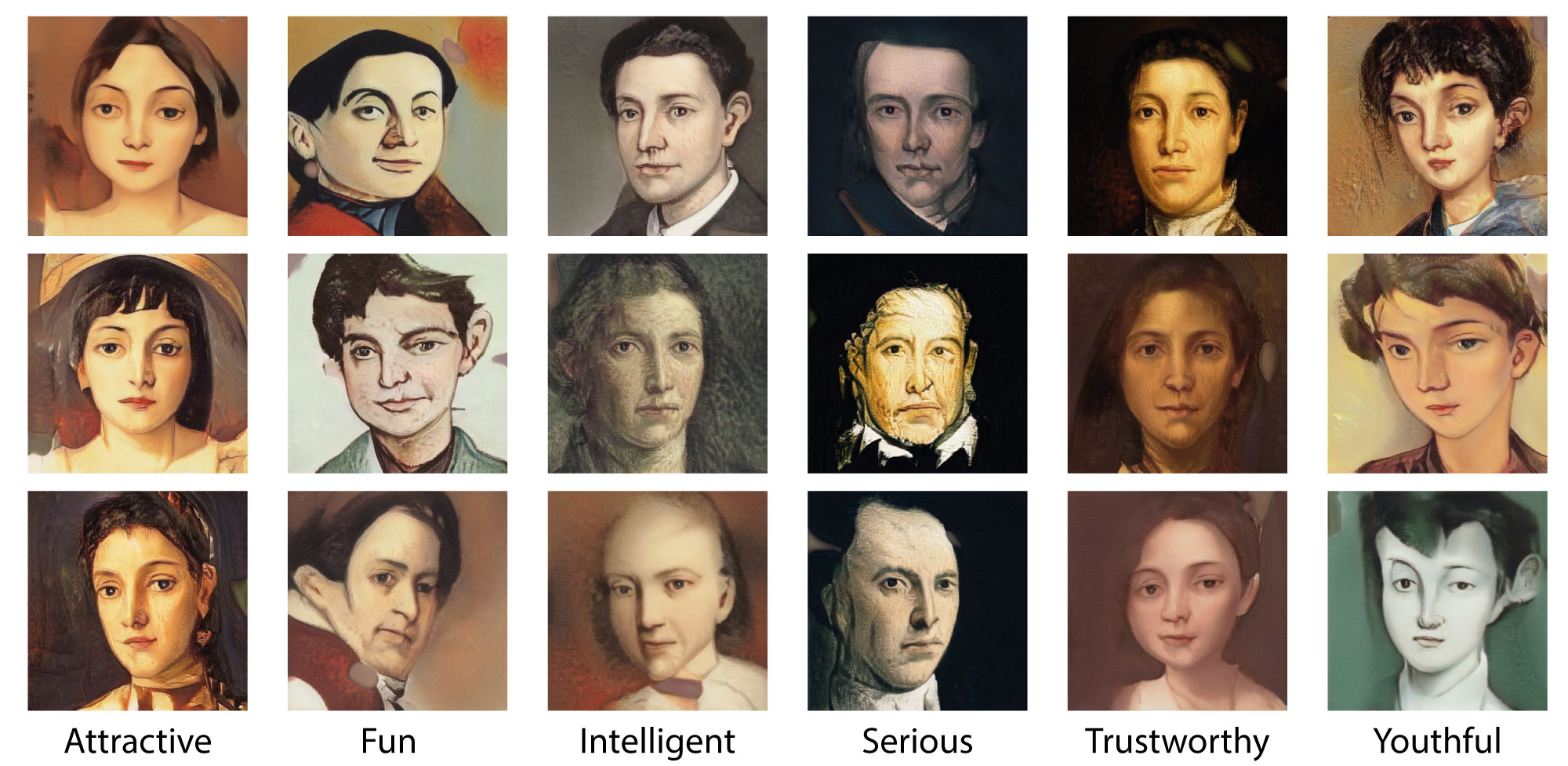}
  \caption{Final GSP samples from Exp.\ 4h, which used the StyleGAN model pretrained on a dataset of portraits from WikiArt (\url{https://www.wikiart.org/}).}
  \label{fig:art-faces}
\end{figure}

\textbf{Training dataset.} To provide a more intuitive illustration of the method's dependence on the training dataset, Fig.\ \ref{fig:art-faces} displays final GSP samples from Exp.\ 4h, which used the StyleGAN model trained on a dataset of portraits from WikiArt (\url{https://www.wikiart.org/}). 
The artistic nature of the WikiArt dataset differs clearly from the photographic nature of the FFHQ dataset, and this is reflected in the GSP samples. 
Nonetheless, the GSP process still successfully navigates this new space to find samples that subjectively reflect the target adjectives.

\subsection{Conclusion}

Our analyses indicate that GSP is an effective tool for exploring the generative space of the StyleGAN model.
Here we relied on a simple PCA approach for creating a reduced basis of the generative space, but there are other promising approaches in the literature that could also be applied to this task (e.g., \citep{voynov2020,shen2020}).
However, our analyses also indicate that dataset bias is a real and important issue when interpreting the outcomes of this approach.
Future work must engage with this problem
by studying the kinds of biases inherent in their generative models
and ideally finding ways to construct less biased models in the first place
(e.g., \citep{grover2019fair}).

\renewcommand{\bibsection}{\section*{Appendix references}}

\putbib
\end{bibunit}

\end{appendices}

\end{document}